\documentclass[11pt]{amsart}
\usepackage[colorlinks = true,
            linkcolor = red,
            urlcolor  = magenta,
            citecolor = black,
            anchorcolor = blue]{hyperref}
\usepackage{color}
\usepackage[alphabetic]{amsrefs}
\usepackage{tikz}
\usepackage{graphicx,amssymb,amsmath,amsthm,amsfonts,mathtext,color,wrapfig,ytableau}
\usetikzlibrary{arrows}
\usepackage[all]{xy}
\usepackage{wrapfig}
\usepackage{graphicx}
\usepackage{epstopdf}
\usepackage{mathrsfs}
\usepackage[vcentermath]{youngtab}
\usepackage[mathcal]{eucal}

\makeatletter \@addtoreset{equation}{section} \makeatother

\renewcommand{\eprint}[1]{\href{https://arxiv.org/abs/#1}{#1}}

\BibSpec{article}{%
    +{}  {\PrintAuthors}                {author}
    +{,} { \textit}                     {title}
    +{.} { }                            {part}
    +{:} { \textit}                     {subtitle}
    +{,} { \PrintContributions}         {contribution}
    +{.} { \PrintPartials}              {partial}
    +{,} { }                            {journal}
    +{}  { \textbf}                     {volume}
    +{}  { \PrintDatePV}                {date}
    +{,} { \issuetext}                  {number}
    +{,} { \eprintpages}                {pages}
    +{,} { }                            {status}
    +{,} { \DOI}                   {doi}
    +{,} { \eprint}        {eprint}
    +{}  { \parenthesize}               {language}
    +{}  { \PrintTranslation}           {translation}
    +{;} { \PrintReprint}               {reprint}
    +{.} { }                            {note}
    +{.} {}                             {transition}
    +{}  {\SentenceSpace \PrintReviews} {review}
}

\def\be{\begin{eqnarray}}
\def\ee{\end{eqnarray}}

\newcommand{\Reals}{\mathbb{R}}

\newcommand{\CN}{\mathcal{N}}

\newcommand{\CZ}{\mathcal{Z}}

\makeatletter
\newcommand{\ellSN}{\mathop{\operator@font sn}\nolimits}
\newcommand{\ellCN}{\mathop{\operator@font cn}\nolimits}
\newcommand{\ellDN}{\mathop{\operator@font dn}\nolimits}
\newcommand{\ellAM}{\mathop{\operator@font am}\nolimits}
\newcommand{\ellK}{\mathop{\smash{\operator@font K}\vphantom{a}}\nolimits}
\newcommand{\ellE}{\mathop{\smash{\operator@font E}\vphantom{a}}\nolimits}
\makeatother

\ifx\genfrac\sdflkaj

\else

\fi

\newcommand{\beq}{\begin{equation}}
\newcommand{\eeq}{\end{equation}}

\newtheorem{theorem}{Theorem}[section]

\newtheorem{proposition}[theorem]{Proposition}

\newtheorem{conjecture}[theorem]{Conjecture}

\makeatletter
\def\mr@ignsp#1 {\ifx\:#1\@empty\else #1\expandafter\mr@ignsp\fi}%
\newcommand{\multiref}[1]{\begingroup
\xdef\mr@no@sparg{\expandafter\mr@ignsp#1 \: }%
\def\mr@comma{}%
\@for\mr@refs:=\mr@no@sparg\do{\mr@comma\def\mr@comma{,}\ref{\mr@refs}}%
\endgroup}
\makeatother

\newcommand{\hypref}[2]{\ifx\href\asklfhas #2\else\href{#1}{#2}\fi}
\newcommand{\Secref}[1]{Section~\multiref{#1}}
\newcommand{\secref}[1]{Sec.~\multiref{#1}}

\newcommand{\appref}[1]{App.~\multiref{#1}}

\newcommand{\figref}[1]{Fig.~\multiref{#1}}
\renewcommand{\eqref}[1]{(\multiref{#1})}

\def\[{\begin{equation}}
\def\]{\end{equation}}
\def\<{\begin{eqnarray}}
\def\>{\end{eqnarray}}

\ifx\href\asklfhas\newcommand{\href}[2]{#2}\fi

\title{The Quantum DELL System}

\author{Peter Koroteev}
\address[Peter Koroteev]{\newline
Department of Mathematics,\newline
University of California Berkeley,\newline
Evans Hall 970,\newline
Berkeley CA 94720,\newline
United States of America\newline
 Email: \href{mailto:pkoroteev@math.berkeley.edu}{pkoroteev@math.berkeley.edu}\newline
 \href{https://www.math.ucdavis.edu/research/profiles/?fac_id=pkoroteev}{website}}

\author{Shamil Shakirov}
\address[Shamil Shakirov]{\newline
Shamil Shakirov,\newline
Department of Mathematics, \newline
Uppsala University,\newline
Box 480 751 06,\newline
Uppsala, Sweden\newline
\href{mailto:shamil.shakirov@math.uu.se}{shamil.shakirov@math.uu.se}\newline
\href{https://katalog.uu.se/profile/?id=N19-572}{website}
}

\begin{document}

\begin{abstract}
We propose quantum Hamiltonians for the double elliptic many-body system (DELL). The Hamiltonians depend elliptically on both the coordinates and the momenta. We conjecture that their common eigenfunction is the normalized Nekrasov--Shatashvili limit of the instanton partition function of the six-dimensional gauge theory with adjoint matter and a codimension-two defect. We test the conjecture in low rank. When the coordinate dependence becomes trigonometric ($p=0$), the Hamiltonians act triangularly on symmetric polynomials and their diagonal entries are explicit. For two particles we identify the eigenpolynomials in this limit with an elliptic generalization of the Macdonald polynomials.
\end{abstract}
\maketitle

\setcounter{tocdepth}{1}
\tableofcontents

\section{Introduction and Main Results}
Integrable many-body systems are deeply interconnected with other branches of modern mathematical physics -- representation theory, algebraic geometry, string and gauge theory. We shall be discussing complex algebraic integrable systems with $N$ degrees of freedom, with a phase space being a Lagrangian fibration of complex dimension $2N$, equipped with a holomorphic symplectic 2-form, over a smooth base whose fibers are Abelian varieties\footnote{See i.e. section 4 of \cite{Nekrasov:2013xda} for review and references}. In proper Darboux coordinates $(\textbf{p},\textbf{q})=(p_1,\dots, p_N;q_1,\dots, q_N)$ the symplectic form reads $\Omega=\sum_{i}d p_i \wedge dq_i$ and there are $N$ Poisson commuting Hamiltonians $H_1(\textbf{p},\textbf{q}),\dots, H_N(\textbf{p},\textbf{q})$. One can study Hamiltonians in action-angle variables so that they only depend on action variables. In these variables the Hamiltonian evolution is linearized on the fibers which serve as the level sets of the Hamiltonians.

Since the discovery of the Seiberg-Witten solution of $\mathcal{N}=2$ supersymmetric gauge theories \cite{Seiberg:1994aj,Seiberg:1994rs} and integrable systems related to them \cite{Gorsky:1995zq} there has been a significant progress in understanding of integrable models and their dualities. The integrable system arises in the infrared description of the above mentioned gauge theories, which are governed by a holomorphic prepotential $\mathcal{F}(\textbf{a})$, where $\textbf{a}=(a_1,\dots a_N)$ are the vacuum expectation values of vector multiplets of the supersymmetric theory whose gauge group has rank $N$. The $N$-dimensional Abelian variety parameterized by the period matrix $\tau_{ij}= \partial\mathcal{F}/\partial a_i\partial a_j$ is fibered over the Coulomb branch of the theory with coordinates $\textbf{a}$.
\begin{center}
\begin{figure}[!h]
\includegraphics[scale=0.5]{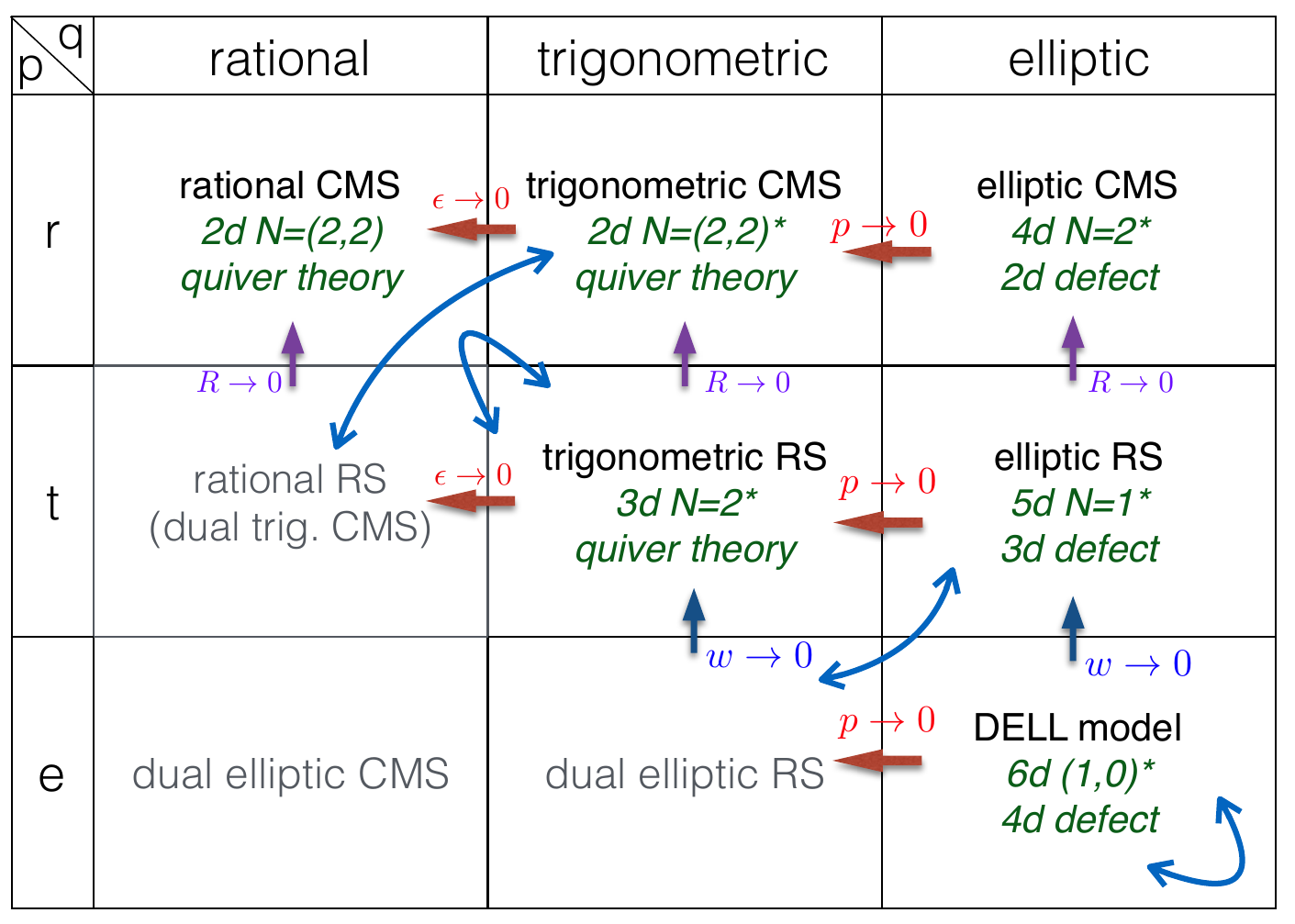}
\caption{The ITEP Table \cite{Mironov:2000if} of integrable many-body systems according to their periodicity properties in coordinates $q$ (columns) and momenta $p$ (rows) together with their gauge theory realizations.}
\label{fig:ITEPtable}
\end{figure}
\end{center}
The Abelian nature of the Lagrangian fibers in the algebraic description of the integrable system suggests that each coordinate and momentum can take values in complex line $\mathbb{C}$, in a complex line without origin $\mathbb{C}^\times$, or in an elliptic curve $\mathcal{E}= \mathbb{C}^\times/q^{\mathbb{Z}}$, where $q$ is a complex parameter. Therefore all many-body integrable models can be classified into \textit{rational, trigonometric} and \textit{elliptic} depending on where its coordinates and momenta take their values. This is summarized in \figref{fig:ITEPtable}.
The first row of the table contains Calogero-Moser-Sutherland (CMS) family, the second row describes Ruijsenaars-Schneider (RS) family, whereas the last row contains models that are bispectrally dual to elliptic CMS, elliptic RS, and, finally, the DELL system. Blue double arrows describe bispectral dualities ($p-q$ dualities or Fourier transforms) between the corresponding models. Color arrows ($w\to 0, R\to 0$, etc.) show the limiting cases when the corresponding model degenerates into a simpler model. The diamond of integrability organizes these nine systems by the two degenerations. In this labeling DELL occupies oval~1, dual elliptic RS occupies oval~2 ($p\to0$), and ordinary elliptic RS occupies oval~3 ($w\to0$). The first author's forthcoming monograph develops the corresponding oper and Bethe descriptions \cite{DiamondPrep}. 

An explicit example of such integrable system was discussed in \cite{Gorsky:1995zq,Martinec:1995by,Donagi:1995cf}. In particular, the $\mathcal{N}=2^*$ theory with gauge group $U(N)$ gives rise to $N$-body elliptic Calogero-Moser model whose coupling constant is related to the mass of adjoint hypermultiplet. In this case the Liouville tori of algebraic integrable systems can be found inside the Jacobians of an algebraic curve, in other words, we are dealing with the Hitchin system \cite{hitchin1987}. According to \cite{Donagi:1995cf}, a point in such integrable system is a holomorphic vector bundle together with and adjoint-valued one form on a genus one Riemann surface $\Sigma$. One can construct a spectral curve inside the cotangent bundle to this Riemann surface $\text{det}(u-\phi)=0$, where $u$ lives in the canonical bundle of the curve $u\in K_\Sigma$. The spectral curve gives rise to Hamiltonians of the integrable system depicted in the top-right corner of the table in \figref{fig:ITEPtable}.

\subsection{String/Gauge Theory Background}
\begin{figure}
\includegraphics[scale=0.8]{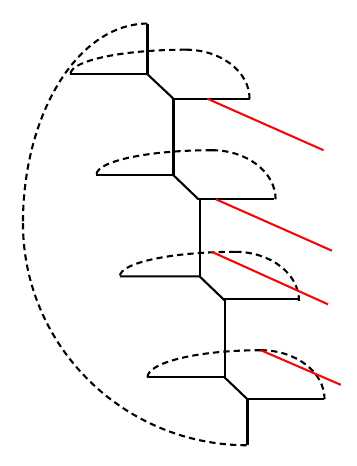}
\caption{Toric diagram for 6d $(1,0)^*$ $U(4)$ theory in the presence of codimension two defect (red branes). The NS5 branes is stretched vertically, the D5 branes are stretched horizontally.}
\label{fig:toricgiag}
\end{figure}

String theory provides us with a toolkit to explore the dualities between gauge theories in various dimensions and between integrable systems which are dual to these gauge theories. In particular, there is a duality between 5d $\CN=1^*$ theories and 6d theories with fundamental matter \cite{Hollowood:2003cv}. Toric diagrams in the geometric description of those theories can be obtained one from another by a 90-degree rotation. Both models may be considered as limiting cases of the 6d theory with adjoint matter compactified on a torus with $(1,0)^*$ supersymmetry, see \figref{fig:toricgiag}.

Let us consider little string theory (LST) (see \cite{WittenLST,Losev:1998,Seiberg:1997} and \cite{Aharony:2000} for review) on the NS5 brane within Type IIA construction with compact direction of radius $R$. Let $\phi$ be the dilaton expectation value. We can apply T-duality along the compact direction. After the T-duality we obtain the NS5 brane in Type IIB theory compactified on a circle of radius
\begin{equation}
R'=\alpha'\frac{e^{-\phi}}{R}\,.
\end{equation}
Now we can apply S-duality in Type IIB theory -- the NS5 brane will turn into D5 brane and the dilaton will change its value to $\phi'=-\phi$. Since we are dealing with the D5 brane now we can identify the field theory which lives on it -- the $U(1)$ 6d $(1,1)$ theory on $\mathbb{R}^4\times S^1_{R'}\times S^1_{R''}$, where $S^1_{R''}$ is an additional spectator compact direction on the D5 brane which was present on all branes from the beginning. This 6d theory has the following coupling constant
\begin{equation}
\frac{1}{g^2_{\text{6d}}}=\frac{e^{-\phi'}}{\alpha'}=\frac{e^{\phi}}{\alpha'}\,.
\end{equation}
When the dilaton VEV is sent to infinity the theory becomes weakly coupled. Equivalently we can view this 6d theory as a 5d $\CN=2$ theory on $\mathbb{R}^4\times S^1_{R''}$ with coupling
\begin{equation}
\frac{1}{g^2_{\text{5d}}}=R'\frac{1}{g^2_{\text{6d}}}=\alpha'\frac{e^{-\phi}}{R}\frac{e^{\phi}}{\alpha'}=\frac{1}{R}\,.
\end{equation}
Note that the 5d coupling constant depends only on the compactification radius\footnote{We thank Francesco Benini for discussions about this topic.}.

The above construction of p-q webs with D5 and NS5 branes can now be generalized to engineer more complex (quiver) 6d $(1,1)$ theories. In addition, as it has been done so in lower dimensions, we shall introduce mass of adjoint hypermultiplets, therefore deforming $(1,1)$ supersymmetry in 6d to $(1,0)^*$ and $\CN=2$ supersymmetry in 5d to $\CN=1^*$.

\subsubsection{$Bispectral/T$-Duality}
One starts with a 6d $(1,0)^*$ theory represented by a p-q web as in \figref{fig:toricgiag}. Let us call this theory A. By rotating the brane web by 90 degrees we arrive to a (generically) different theory, with Neveu-Schwarz and Dirichlet directions of its brane web interchanged. Let's call it theory B. Recall that this 90 degree rotation is described by T-duality with respect to the compact circle inside the worldvolume of the NS branes. In addition to that in each p-q web Dirichlet fivebranes are also compact and the radius of the corresponding circle is $R$. Thus the T-duality merely exchanges $R$ and $R'$. This corresponds to Fourier-Mukai transform at the level of the integrable system which is depicted by diagonal arrows in \figref{fig:ITEPtable}).

\begin{figure}[!h]
\includegraphics[scale=0.8]{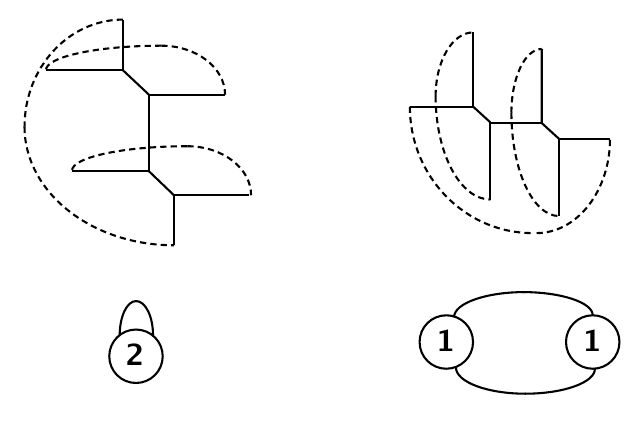}
\caption{A Pair of dual 6d theories (bottom) and their brane realizations (top).}
\label{fig:6dEX}
\end{figure}

Let us look at a simple example. Consider rank two little string theory corresponding, by means of the argument we have just presented, to $\widehat{A}_1$ 6d $(1,0)^*$ theory. In this case theory A is 5d $\mathcal{N}=1^*$ $U(2)$ gauge theory on $\mathbb{R}^4\times S^1$, while theory B is the 6d affine quiver gauge theory with two $U(1)$ gauge groups connected to each other (see \figref{fig:6dEX}). 
Note that in circular quivers an overall $U(1)$ factor inside the gauge group of the whole theory decouples and can be gauged away. This factor can be identified with one of the $U(1)$ groups in this example. The remaining $U(1)$ gauge group has two fundamental hypermultiplets which are rotated by global $SU(2)$ symmetry. In other words, theory B is nothing but $U(1)$ 6d theory with two flavors as shown in \figref{fig:6dEX}
\begin{figure}[!h]
\includegraphics[scale=0.8]{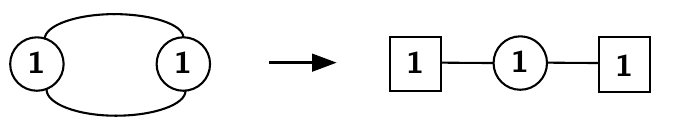}
\caption{Decoupling the $U(1)$ gauge factor in theory B from Fig \ref{fig:6dEX}.}
\label{fig:6d12split}
\end{figure}
Note that if the ranks of the individual gauge groups of theory B above at each node are higher the quiver remains affine. This will happen when theory A will be a quiver gauge theory with several nodes. The weakly coupled regime of the above theories can be accessed by sending $R'\to 0$ keeping $R$ fixed, or equivalently, we can send the dilaton VEV to infinity.

\subsection{Classical DELL System}
In this paper we shall discuss canonical quantization of the DELL model which was studied in \cite{Braden:1999aj,Mironov:1999vj,Braden:2001yc,Hollowood:2003cv}. As it was pointed out in \cite{Hollowood:2003cv} term `double elliptic' is a misnomer, as its momenta and coordinates as complex two-vectors lie in a 4-torus. Nevertheless, by tradition and for brevity, we shall refer to this model as DELL thereby emphasizing its double periodic nature. More recently DELL systems were studied from the point of view of modular transformations \cite{Aminov:2013asa,Aminov:2014wra,Aminov:2017dud}.

In \cite{Braden:2003gv,Hollowood:2003cv} it was shown how to construct a spectral curve (equivalently, Seiberg-Witten curve of the 6d theory on $\mathbb{R}^4\times T^2$) of the $N$-body DELL model using the approach of matrix models and M-theory. The description involves line bundles over polarized abelian variety whose sections are generalized (higher genus) theta-functions. The smooth Seiberg--Witten curve is a divisor of polarization type $(1,N)$ on an abelian surface. Its self-intersection is $2N$, so adjunction gives genus $N+1$.
The Seiberg-Witten (DELL spectral) curve can be represented in the following explicit form (see \cite{Braden:2003gv} for the detailed description)
\begin{equation}
\sum_{n=0}^{\infty} \frac{m^n}{n!} \left[\partial_\xi^n \vartheta (e^\xi|p)\right]\cdot\left[\partial_\zeta^n H(\zeta,w,\textbf{a})\right]=0\,,
\label{eq:gN2curve}
\end{equation}
where
\begin{equation}
H(\zeta,w,\textbf{a})=\prod_{i=1}^N\vartheta\left(e^\zeta/a_i\mid w^2\right)\,,
\end{equation}
Here $m=\log t$ is the dimensionless mass in the logarithmic coordinates $\xi,\zeta$, and $\vartheta$ is the odd theta function in \eqref{eq:ShiftedTheta}. The factorial comes from expanding $\exp(m\partial_\xi\partial_\zeta)$ acting on $\vartheta(e^\xi\mid p)H(\zeta,w,\textbf a)$, as in Eq.~(2.20) of \cite{Braden:2003gv}. In the cycle basis of that paper the two pullback differentials are
\begin{equation}
\frac{d\xi}{2\pi i}=\sum_{i=1}^N\omega_i\,,\qquad \frac{d\zeta}{2\pi i}=\omega_{N+1}\,.
\end{equation}
The normalized holomorphic differentials and the full period matrix on $\Sigma$ are specified by the $N+1$ pairs of cycles $\{A_i,B_i\}_{i=1}^{N+1}$,
\begin{equation}
\oint\limits_{A_i} \omega_j = \delta_{ij}\,,\qquad \oint\limits_{B_i} \omega_j = \Pi_{ij}\,,
\end{equation}
where $\sum_{j=1}^N \Pi_{ij}=\tau_{\mathrm{YM}}$ for $1\leq i\leq N$. For points $P_0,P_1,\dots,P_{N-1}$ on $\Sigma$, the Abel map has components
\begin{equation}
x_i = \sum_{j=1}^{N-1}\int\limits_{P_0}^{P_j} \omega_i\,,
\label{eq:coordinatesAbel}
\end{equation}
Projecting this class to the reduced Jacobian $\mathcal{J}_0(\Sigma)$ gives the $N-1$ relative angle variables \cite{Braden:2003gv}.


\subsection{Quantization}
As it was first shown in \cite{Bullimore:2015fr} and later proven by Nekrasov (in the case of the four-dimensional $U(2)$ theory) \cite{Nekrasov:2009rc,Alday:2010vg, Nekrasov:2013xda,Nekrasov:2015wsu,Nekrasov:2017gzb} how to find a formal spectrum of quantum elliptic integrable systems of Calogero or Ruijsenaars type. The eigenfunctions of the corresponding Hamiltonians were shown to be supersymmetric partition functions in the presence of monodromy-type defects
of the Gauge/Bethe dual 4d and 5d theories with adjoint matter respectively. The eigenfunctions were represented by vacuum expectation values of local chiral observables (in 4d) and Wilson lines (in 5d):
\begin{equation}
H_i \mathcal{Z}(p, \textbf{x})= E_i(p,\textbf{a})\mathcal{Z}(p,\textbf{x})\,,
\label{eq:HamEigen}
\end{equation}
where both $\mathcal{Z}(p, \textbf{x})$ and $E_i(p,\textbf{a})$ are series expansions in the instanton counting parameter $p$.
As was argued in \cite{Alday:2009fs} and used later in \cite{Gaiotto:2013sma} the positions of a surface operator in the $U(N)$ $\mathcal{N}=2$ gauge theory are given by integrals over open paths of the corresponding K\"ahler classes $\omega_i$ \eqref{eq:coordinatesAbel}. This observation explains why insertions of defect operators should be treated as the coordinates of the (complexified) integrable systems and the wavefunction $\mathcal{Z}(p, \textbf{x})$.

Mathematically the eigenfunctions of the above Hamiltonians are equivariant integrals of certain characteristic classes over the affine Laumon spaces. Negu\c{t} treated the Calogero--Sutherland system on ordinary Laumon spaces \cite{Negut_2009} and the elliptic Calogero--Moser system on affine Laumon spaces \cite{NegutAffine}. In \cite{Koroteev:2018isw} a conjecture was made stating that the K-theoretic equivariant Euler characteristic of the affine Laumon space is the eigenfunction of the elliptic Ruijsenaars-Schneider model.

\subsection{Main Results}
We propose the new quantum integrable system, called DELL, which for $N-1$ degrees of freedom\footnote{$N$ degrees of freedom with removed center of mass} is described by Hamiltonians
\begin{equation}
\mathcal{\widehat{H}}_a = \widehat{\mathcal{O}}_0^{-1} \widehat{\mathcal{O}}_a\,, \qquad a = 1, \ldots, N-1\,,
\label{eq:DELLHAMs}
\end{equation}
where operators $\mathcal{O}_0,\mathcal{O}_1,\dots,\mathcal{O}_{N-1}$ are Fourier modes of the following current
\begin{multline}
\widehat{\mathcal{O}}(z) = \sum_{n \in {\mathbb Z}} \widehat{\mathcal{O}}_n\, z^n \\
= \sum_{n_1, \ldots, n_N\in\mathbb{Z}} (-z)^{\sum n_i}\, w^{\sum \frac{n_i(n_i - 1)}{2}} \prod_{i < j} \vartheta\big( t^{n_i - n_j} \widehat{ x}_i / \widehat{ x}_j\vert p \big)\, \widehat{ p}_1^{n_1} \cdots \widehat{ p}_N^{n_N}\,.
\label{eq:Ocurrent}
\end{multline}
Here $\vartheta$ is the odd theta function \eqref{eq:ShiftedTheta}, and the canonically conjugate position and momentum operators obeying the canonical $q$-commutation relation $\widehat{p}_j \widehat{x}_i=q^{\delta_{ij}}\widehat{x}_i\widehat{p}_j$ act on functions of positions as
\begin{equation}
\widehat{x}_i f(\textbf{x}) = x_i f(\textbf{x})\,, \qquad \widehat{p}_i f(\textbf{x}) = f(x_1, \ldots, q x_i, \ldots, x_N)\,.
\end{equation}

We shall now formulate two conjectures and one proposition about the properties of the DELL model, which we have checked in a number of cases. We expect that these conjectures will be proven in the near future using methods of enumerative algebraic geometry and geometric representation theory.

\begin{conjecture}
The quantum DELL Hamiltonians $\mathcal{H}_1,\dots \mathcal{H}_{N-1}$ in \eqref{eq:DELLHAMs} commute,
\begin{equation}
[\mathcal{\widehat{H}}_a,\mathcal{\widehat{H}}_b]=0\,,\qquad a\neq b\,.
\end{equation}
\label{Th:ConjComm}
\end{conjecture}
We have checked the commutativity of the DELL Hamiltonians for $N=3$ through order $w^5$ and for $N=4$ through order $w^3$ in the expansion in $w$, using unexpanded theta functions at numerical values of $p,q,t$ with 40-digit precision, and in exact arithmetic at $p=0$ on spaces of symmetric polynomials of low degree (Section~\ref{sec:p0exact}). For $N=2$ there is a single Hamiltonian $\widehat{\mathcal{H}}_1$ besides the central element $\widehat{\mathcal{H}}_2=\widehat{p}_1\widehat{p}_2$, so there is nothing to check. Simple deformations of the current, for instance replacing $t^{n_i-n_j}$ by $t^{\,\mathrm{sign}(n_i-n_j)}$, destroy commutativity already at order $w$. These checks give us confidence that the conjecture holds. As the reader can see the DELL Hamiltonians are highly non-local as they involve formal infinite series of functions of shift operators, which severely limits the computational ability to verify the conjecture. We expect that the general proof to be found some time in the near future.
\smallskip\\

The DELL Hamiltonians provide a two-parameter $(w, p)$ generalization of the trigonometric Ruijsenaars-Schneider integrable system, whose eigenfunctions are also known as Macdonald polynomials. The common eigenfunctions of the full DELL Hamiltonians are therefore a natural candidate to replace the Macdonald polynomials in the double-elliptic setting, and we will call them double elliptic Macdonald functions. They depend on four parameters -- $q,t$ and $p,w$. We expect them to agree with `\textit{affine Macdonald polynomials}' in representation theory \cite{Etingof:ab}, similar in the number of parameters and limiting behavior.

We propose a description of the formal spectrum in terms of defect partition functions of a 6d gauge theory on $\mathbb{R}^4\times T^2$.
\begin{conjecture}\label{th:Conj2}
Let $\textbf{x}=(x_1,\dots,x_N)$ be the position vector, $\mathscr{Z}(p,\textbf{x})$ be a properly normalized equivariant elliptic genus of the affine Laumon space in the limit $\epsilon_2\to 0$. Then there exists a function $\lambda(z,\textbf{a},w,p)$ such that
\begin{equation}
\widehat{\mathcal{O}}(z)\mathscr{Z}(p,\textbf{x}) = \lambda(z, \textbf{a},w,p) \ \widehat{\mathcal{O}}_{0} \mathscr{Z}(p,\textbf{x})\,.
\label{eq:Oeignproblem}
\end{equation}
In particular, by expanding currents $\widehat{\mathcal{O}}(z)$ as in \eqref{eq:Ocurrent} and $\lambda(z,\textbf{a},w,p)=\sum_n \lambda_n(\textbf{a},w,p) z^n $ in $z$ we obtain similar relations for each operator $\widehat{\mathcal{O}}_n$, or, using \eqref{eq:DELLHAMs} we obtain the eigenvalue problem for DELL  Hamiltonians
\begin{equation}
\widehat{\mathcal{H}}_n \mathscr{Z}(p,\textbf{x}) = \lambda_n(\textbf{a},w,p)\, \mathscr{Z}(p,\textbf{x})\,,\qquad \lambda_0=1\,.
\end{equation}
\end{conjecture}

At the moment we know the eigenvalues $\lambda_n(\textbf{a},w,p)$ in terms of first several terms in series expansion in $w$ and $p$. In the limiting cases, when either $w$ or $p$ vanish (and we land onto the eRS model or on the dual RS model) we have complete control of these functions (for $p=0$ see Section~\ref{sec:p0exact}) and know of their interpretations in terms of the corresponding gauge theories. Yet, at the moment we are lacking a complete understanding of a geometric or physical (in terms of a vacuum expectation value of some BPS observable in the 6d theory) meaning of $
\lambda_n(\textbf{a},w,p)$. We shall however make an educated guess about these matters in \secref{Sec:GuessEigen}. We hope to understand it in the nearest future.

However, in the limit $w\to 0$, in which the DELL model becomes the eRS model, the eigenvalues can be expressed through the eigenfunction itself. The following result is inspired by the study of Chern-Simons theory on $S^3$ and complements the result of \cite{Bullimore:2015fr}:
\begin{proposition}\label{Conj1}
Let $\textbf{x}=(x_1,\dots x_N)$ be the position vector of the eRS model and let $\mathcal{Z}^{{\text{RS}}}(\textbf{a},\textbf{x})=\lim\limits_{w\to 0}\mathscr{Z}(p,\textbf{x})$ be a common eigenfunction of the eRS Hamiltonians $\widehat{\mathcal{H}}_i\big|_{w=0}$,
\begin{equation}
\widehat{\mathcal{H}}_i \mathcal{Z}^{{\text{RS}}}(\textbf{a},\textbf{x}) = \lambda_i (\textbf{a})\mathcal{Z}^{{\text{RS}}}(\textbf{a},\textbf{x})\,,\qquad i=1,\dots,N-1\,,
\end{equation}
which is regular at the points $\textbf{x}=t^{\vec\rho}q^{\vec e_I}$ for all $I\subset\{1,\dots,N\}$, where $\vec e_I$ is the indicator vector of $I$, and satisfies $\mathcal{Z}^{\text{RS}}(\textbf{a},t^{\vec\rho})\neq 0$. Then, for generic $t$, the eigenvalues read
\begin{equation}
\lambda_i(\textbf{a})=(-1)^i\prod_{k=1}^{i}\frac{\vartheta(t^{N-k+1}|p)}{\vartheta(t^{k}|p)}\;\frac{\mathcal{Z}^{{\text{RS}}}(\textbf{a},t^{\vec{\rho}}q^{\vec{\omega}_i})}{\mathcal{Z}^{{\text{RS}}}(\textbf{a},t^{\vec{\rho}})}\,,\qquad i=1,\dots, N-1\,,
\end{equation}
where $\rho_k=\frac{N+1}{2}-k$ are the components of the $SU(N)$ Weyl vector $\vec{\rho}$, we use the $GL(N)$ lift of the $i$-th fundamental weight $\vec{\omega}_i$ with components $(\omega_i)_k=1$ for $k\leq i$ and $0$ for $k>i$, and $t^{\vec\rho}q^{\vec\omega_i}$ stands for $x_k=t^{\rho_k}q^{(\omega_i)_k}$.
\end{proposition}
\begin{proof}
At $w=0$ only $n_k\in\{0,1\}$ contribute to \eqref{eq:Ocurrent}, and
\begin{equation*}
\widehat{\mathcal{H}}_i=(-1)^i\sum_{|I|=i}\ \prod_{k\in I,\,l\notin I}\frac{\vartheta(t\, x_k/x_l|p)}{\vartheta(x_k/x_l|p)}\ \prod_{k\in I}\widehat{p}_k\,.
\end{equation*}
At $x_k=t^{\rho_k}$ we have $x_k/x_l=t^{l-k}$. If $I$ is not the initial segment $\{1,\dots,i\}$, it contains some $k$ with $k-1\notin I$, and the factor with $l=k-1$ is $\vartheta(1|p)=0$; it multiplies $\mathcal{Z}^{\text{RS}}(\textbf{a},t^{\vec\rho}q^{\vec e_I})$, which is finite by assumption. For $I=\{1,\dots,i\}$ the product telescopes to $\prod_{k=1}^{i}\vartheta(t^{N-k+1}|p)/\vartheta(t^{k}|p)$. Evaluating the eigenvalue equation at $\textbf{x}=t^{\vec\rho}$ gives the claim.
\end{proof}
For $i=1$ and $i=N-1$ the theta-product in the prefactor reduces to $\vartheta(t^N|p)/\vartheta(t|p)$. We have also confirmed the formula on the perturbative eigenfunctions for $N=3$ through order $p^2$ and for $N=4$ through order $p$.
This gives the elliptic RS eigenvalue at $w=0$ in terms of two values of its eigenfunction. An evaluation formula in this limit also appears in \cite{Gorsky:2022aa}. This generalizes the well-known nonelliptic result, and may have topological implications in terms of the elliptic Chern-Simons theory, which remains to be constructed.

\subsection{Connection to Aminov-Mironov-Morozov DELL Hamiltonians}
There is a different approach to the DELL system in the literature which is discussed in \cite{Mironov:1999vi} and more recently in \cite{Aminov:2016ruk}. The explicit $N$-body Hamiltonians are written in formula (2.1) of \cite{Aminov:2016ruk} and possess some similarity with ours, however, characteristics \textbf{a} and \textbf{b} in the definition of theta functions \eqref{eq:ThetaDef} are slightly different. More importantly, theta functions which we use in our paper (they are borrowed from \cite{Braden:2003gv}) have constant period matrix \eqref{eq:PeriodMat}, whereas the period matrix from the Aminov-Mironov-Morozov paper $T_{ij}=\partial_i \partial_j \mathcal{F}$ is the hessian of the 6d Seiberg-Witten prepotential which is known by the authors up to several orders in instanton coupling. A comparison of the curves must also identify the abelian-surface factor and the reduced Jacobian in the two descriptions.

It is clear at this point that a significant amount of effort is needed in order to match the two constructions; nevertheless we stay optimistic in this regard given the fact that both models arise from the same gauge theory.\footnote{We thank the authors of \cite{Aminov:2016ruk} for continuing discussions on this.}

\subsection{Future Directions}
Several questions about the quantum DELL system can be approached through its degenerations. 

\begin{itemize}
\item F-theory constructions give Seiberg--Witten curves for six-dimensional quiver theories of several simply laced types \cite{Haghighat:2016jjf,Haghighat:2018dwe}. It would be interesting to quantize these curves and determine the interactions between adjacent quiver nodes.

\item A representation-theoretic construction of the current \eqref{eq:Ocurrent} could explain why its zero mode $\widehat{\mathcal O}_0$ normalizes the Hamiltonians in \eqref{eq:Oeignproblem}. The Sklyanin algebra and elliptic double affine Hecke algebras are candidates to investigate. Such a construction might also provide a way to prove the commutativity and defect-eigenfunction conjectures of Section~1.4.

\item Inozemtsev scaling relates elliptic Calogero--Moser to affine Toda and elliptic Ruijsenaars--Schneider to affine $q$-Toda \cite{inozemtsev1989}. For quantum DELL, double Inozemtsev limits scale both elliptic parameters together with the coupling \cite{Gorsky:2022aa}. On the gauge-theory side this decouples the massive adjoint hypermultiplet. One can ask how this scaling acts on the defect eigenvalue equation \eqref{eq:Oeignproblem}. See also \cite{Aminov:2017dud}.

\item At large $N$ the elliptic Ruijsenaars--Schneider spectrum yields the spectrum of the difference quantum intermediate long wave system \cite{Koroteev:2016,Koroteev:2018a}. Its energies are related to quantum multiplication by tautological classes in the equivariant $K$-theory of the $U(1)$ instanton moduli space. What geometric operation accounts for the second elliptic parameter in the corresponding DELL limit?

\item The first author's forthcoming monograph \emph{The Diamond of Integrability} \cite{DiamondPrep} studies the oper and Wronskian descriptions of the nine systems in \figref{fig:ITEPtable}. On the marked chart for the DELL corner the Wronskian determines the classical Hamiltonian current and gives geometric Bethe equations. A separate problem is to construct a quantum lattice Hamiltonian with those equations and to identify its limits at the ordinary and dual elliptic Ruijsenaars--Schneider corners.

\end{itemize}

\subsection{Structure of the Paper}
In the next Section we remind the reader about the classical Seiberg-Witten geometry of the 6d theory with adjoint matter with eight supercharges compactified on $T^2$. In \Secref{Sec:QuantumDell} the quantum DELL Hamiltonians are discussed as well as their limits to the eRS and dual eRS models, and the spectrum at $p=0$ is found exactly. \Secref{sec:States} introduces instanton partition functions of the 6d theory with monodromy defects using characters of the affine Laumon space. We test the proposed DELL eigenvalue equation for its normalized Nekrasov--Shatashvili limit. Finally, in \Secref{Sec:EllipticMacPoly} we discuss in more details polynomial solutions of the DELL difference equation. In particular, we address the properties of the new elliptic Macdonald functions.

\subsection{Acknowledgements}
This manuscript took some time to complete primarily due to the technical computational difficulties of expressions  involving double-periodic functions. We would like to thank numerous people with whom we discussed these and other matters, as well as various institutions which we visited in the last several years, but our special acknowledgements go to Babak Haghigat, Wenbin Yan, Can Kozcaz and Francesco Benini who participated in earlier stages of the project.

\section{Seiberg-Witten Geometry}\label{Sec:ClassicalDell}
The construction of \cite{Braden:2003gv,Hollowood:2003cv} uses the M-theory compactification where two distinct tori are involved -- one spatial $T^2$, which is a part of the gauge theory worldvolume, while the other (sometimes referred to as \textit{electro-magnetic}) torus, after the compactification provides the gauge theory with maximal supersymmetry. This supersymmetry is broken to a $(1,0)^*$ subgroup once we include an equivariant action of $\mathbb{C}^\times$ along the remaining flat noncompact directions whose character is equal to $t$ such that $\log t \sim m$, where $m$ is the mass of the adjoint hypermultiplet.

The Seiberg-Witten curve $\Sigma$ for 6d $(1,0)^*$ theory compactified on $T^2_{w}$ was derived in \cite{Braden:2003gv}. The description involves line bundles over polarized abelian varieties whose sections are generalized theta-functions. Such genus-$g$ theta function with period $g\times g$-matrix $\Pi$ reads
\begin{equation}
\Theta\left[\begin{array}{c}
\textbf{a} \\
\textbf{b}
\end{array}\right]  (\textbf{Z}\vert \Pi)=\sum\limits_{\textbf{m}\in\mathbb{Z}^g}e^{\pi i (\textbf{m}+\textbf{a})\cdot\Pi\cdot(\textbf{m}+\textbf{a}) + 2\pi i (\textbf{Z}+\textbf{b})\cdot(\textbf{m}+\textbf{a})}\,,
\label{eq:ThetaDef}
\end{equation}
where $\textbf{a},\textbf{b},\textbf{Z}$ and $\textbf{m}$ are $g$-dimensional vectors.

The Seiberg-Witten curve can be written in terms of genus-$(N+1)$ theta function as follows
\begin{equation}
\Theta\left[\begin{array}{ccc}
\frac{1}{2} & \cdots & \frac{1}{2}  \\
\frac{1}{2} & \cdots & \frac{1}{2}
\end{array}\right]  \left(\frac{\xi}{2\pi i},\frac{\zeta-\log a_1}{2\pi i},\dots,\frac{\zeta-\log a_N}{2\pi i}\,\middle|\,\widehat{\Pi}\right)=0\,,
\end{equation}
using the following $(N+1)\times(N+1)$ period matrix
\begin{equation}
\widehat{\Pi} =
\begin{pmatrix}
\tau_1& \frac{m}{2\pi i} & \frac{m}{2\pi i}& \cdots & \frac{m}{2\pi i}\\
\frac{m}{2\pi i}& \tau_2 & 0 &\cdots & 0\\
\frac{m}{2\pi i}& 0& \tau_2 &\cdots & 0\\
\vdots&\vdots&\vdots &\ddots &0\\
\frac{m}{2\pi i}& 0&\cdots &0&\tau_2
\end{pmatrix}\,,
\label{eq:PeriodMat}
\end{equation}
where $p=e^{2\pi i\tau_1}$ and $w^2=e^{2\pi i\tau_2}$. The off-diagonal entries give exactly the mixed Fourier factor obtained from $\exp(m\partial_\xi\partial_\zeta)$ in \eqref{eq:gN2curve}. The matrix $\widehat{\Pi}$ is the auxiliary theta matrix in this presentation. The periods of the spectral curve itself vary with its moduli.

At $m=0$, Eq.~\eqref{eq:gN2curve} reduces to $\vartheta(e^\xi\mid p)H(\zeta,w,\textbf a)=0$. For distinct $a_i$ modulo $w^{2\mathbb Z}$, one elliptic component meets the other $N$ components in $N$ ordinary nodes. The normalization is the disjoint union of these $N+1$ elliptic curves.

In the genus-two presentation of \cite{Braden:2003gv,Hollowood:2003cv}, after translating the coordinate origins, the same curve is the zero divisor of a section of type $(1,N)$,
\begin{equation}
\sum_{j=0}^{N-1}a_j\,
\Theta\left[\begin{array}{cc} 0 & j/N \\ 0 & 0 \end{array}\right]
\left(\frac{\xi}{2\pi i},\frac{N\zeta}{2\pi i}\,\middle|\,
\begin{matrix}\tau_1 & Nm/(2\pi i)\\ Nm/(2\pi i)&N\tau_2\end{matrix}\right)=0\,.
\end{equation}
Here the $a_j$ in this last display denote the $N$ theta-basis coefficients, defined up to common scale and determined by the zeros of $H$ in the chosen frame. Braden and Hollowood also discuss the two elliptic RS limits \cite{Braden:2003gv}.

On a cyclic cover of the abelian surface, a marked row of theta sections and its translates define a full flag on the regular locus. The corresponding Wronskian gives, after normalization, the classical DELL Hamiltonian current. The first author's forthcoming monograph follows this construction to dual elliptic RS at $p=0$ and ordinary elliptic RS at $w=0$ \cite{DiamondPrep}.

\section{The Quantum DELL Integrable System}\label{Sec:QuantumDell}
In this section we introduce quantum DELL system as a family of formal $q$-difference operators whose commutativity is the subject of Conjecture~\ref{Th:ConjComm}.
In terms of the canonically conjugate positions and momenta these difference operators act as
\begin{equation}
\widehat{x}_i f(\textbf{x}) = x_i f(\textbf{x})\,, \qquad \widehat{p}_i f(\textbf{x}) = f(x_1, \ldots, q x_i, \ldots, x_N)
\end{equation}
Let $\widehat{\mathcal{O}}(z)$ be the DELL current which is a series in formal variable $z$ 
\smallskip
\begin{multline}
\widehat{ \mathcal{O}}(z) = \sum_{n \in {\mathbb Z}} \widehat{ \mathcal{O}}_n\, z^n \\
= \sum_{n_1, \ldots, n_N\in\mathbb{Z}} (-z)^{\sum n_i}\, w^{\sum \frac{n_i(n_i - 1)}{2}} \prod_{i < j} \vartheta\big( t^{n_i - n_j} \widehat{ x}_i / \widehat{ x}_j\vert p \big)\, \widehat{ p}_1^{n_1} \cdots \widehat{ p}_N^{n_N}\,,
\label{eq:CurrentOz}
\end{multline}
\smallskip
where we use the odd genus one theta function\footnote{In terms of the Jacobi theta function, $\vartheta(x|p)=-i\,p^{-1/8}\,\theta_1(u|\tau)$ for $x=e^{2\pi i u}$ and $p=e^{2\pi i\tau}$. The prefactor $x^{-1/2}$ does not cancel in the ratios appearing below, $\vartheta(t^m y|p)/\vartheta(y|p)=t^{-m/2}\,\theta(t^m y|p)/\theta(y|p)$; with it the current \eqref{eq:CurrentOz} changes by the sign of a permutation of the particles, so that the Hamiltonians are symmetric.}
\begin{equation}
\begin{split}
\vartheta(x|p)&=x^{-1/2}\,\theta(x|p)\,,\\
\theta(x|p) &= \prod_{k = 0}^{\infty} (1 - p^{k+1})(1 - p^{k} x)(1 - p^{k+1}/x) = \sum_{k \in\mathbb{Z}} (-x)^k \, p^{\frac{k(k-1)}{2}}\,.
\end{split}
\label{eq:ShiftedTheta}
\end{equation}
It satisfies $\vartheta(1/x|p)=-\vartheta(x|p)$ and $\vartheta(px|p)=-p^{-1/2}x^{-1}\vartheta(x|p)$. We use $\theta$ for the momentum factors and the elliptic Pochhammer symbols, and $\vartheta$ for the individual localization weights in Section~\ref{sec:States}.
Its $z$-modes $\widehat{\mathcal{O}}_n$ are used to construct DELL Hamiltonians.

We have presented DELL Hamiltonians in the form \eqref{eq:CurrentOz}, where the sum is taken over quadratic powers of the elliptic parameter $w$ multiplied by theta functions involving $p$, since it is convenient for the calculations of the spectrum in the next section. Equivalently one could present \eqref{eq:CurrentOz} in a reciprocal form, where the roles of $w$ and $p$ are swapped. 

Despite the fact that formula \eqref{eq:CurrentOz} provides an infinite set of operators $\widehat{\mathcal{O}}_n$, only the first $N$ of them are independent. We shall discuss periodicity in index $n$ momentarily.

\subsection{Quasiperiodicity}
It is easy to show that $\widehat{ \mathcal{O}}(z)$ is quasiperiodic in $z$ with weight
\begin{equation}
\widehat{ \mathcal{O}}(w z) = (-1)^N z^{-N}\, \widehat{P}^{-1}\, \widehat{ \mathcal{O}}(z)\,, \qquad \widehat{P} = \widehat{p}_1 \cdots \widehat{p}_N\,,
\label{eq:Oquasiperiod}
\end{equation}
where the total momentum $\widehat{P}$ commutes with all the modes, since the coefficients in \eqref{eq:CurrentOz} depend only on the ratios $x_i/x_j$; equivalently, $\widehat{\mathcal{O}}_{a+N}=(-1)^N w^{a}\,\widehat{\mathcal{O}}_a\widehat{P}$. Hence $\widehat{ \mathcal{O}}(z)$ can be decomposed in the basis of theta functions of the appropriate level
\begin{equation}
\widehat{ \mathcal{O}}(z) = \sum\limits_{a = 0}^{N-1} \ \widehat{ \mathcal{ O}}_a \ \theta^{(N)}_a(z)\,,
\end{equation}
where the basis consists of the following $N$ functions
\begin{equation}
\theta^{(N)}_a(z) = z^a \ \sum\limits_{n \in {\mathbb Z}} \ w^{N n(n-1)/2 + a n} \ \big( (-z)^N \widehat{P} \big)^{n}\,,
\label{eq:thetaa}
\end{equation}
and represents an elliptic deformation of the monomial basis $\{1, z, \ldots, z^{N-1}\}$ in the space of polynomials of degree $(N-1)$. This implies that $\widehat{ \mathcal{O}}(z)$ contains as much information as the first $N$ modes (from $0$th to $(N-1)$st).

\subsection{DELL Hamiltonians}
We define quantum Hamiltonians of the DELL integrable system as the following $N-1$ operators
\smallskip
\begin{equation}
\mathcal{ \widehat{ H}}_a = \widehat{\mathcal{O}}_0^{-1} \widehat{ \mathcal{O}}_a, \qquad \qquad a = 1, \ldots, N-1\,,
\label{eq:DefDellHam}
\end{equation}
\smallskip
which are manifestly nonlocal. The modes $\widehat{\mathcal{O}}_a$ can fail to commute. Conjecture~\ref{Th:ConjComm} concerns the normalized Hamiltonians in \eqref{eq:DefDellHam}. The numerical and perturbative checks are described in Section~1.4.

One can see from \eqref{eq:CurrentOz} and \eqref{eq:DefDellHam} that expression
\begin{equation}
\mathcal{L}(z):= \widehat{\mathcal{O}}_0^{-1} \widehat{\mathcal{O}}(z) = \sum\limits_{n\in\mathbb{Z}} z^n\, \mathcal{\widehat{H}}_n = \sum\limits_{a=0}^{N-1}\mathcal{\widehat{H}}_a\,\theta^{(N)}_a(z)\,,\qquad \mathcal{\widehat{H}}_0=1\,,
\end{equation}
generates all DELL Hamiltonians and is the elliptic analogue of the characteristic polynomial of a Lax matrix det$(1-z L)$. Indeed, due to quasiperiodicity \eqref{eq:Oquasiperiod} only $\widehat{\mathcal{H}}_1,\dots,\widehat{\mathcal{H}}_{N-1}$ and $\widehat{P}$ are independent, $\widehat{\mathcal{H}}_{a+N}=(-1)^N w^a\,\widehat{\mathcal{H}}_a\widehat{P}$. The modes with negative index do not vanish, e.g. $\widehat{\mathcal{H}}_{-1}=(-1)^N w\,\widehat{\mathcal{H}}_{N-1}\widehat{P}^{-1}$. Thus $z$ plays the role of the spectral parameter.

The commutativity conjecture for the DELL Hamiltonians can thus be written as
\begin{equation}
[\mathcal{L}(z),\mathcal{L}(z')]=0\,.
\end{equation}

\subsection{Example: Rank 1 System}
Let us consider the DELL model which comes from the SU(2) 6d gauge theory in more details. It has been previously understood in \cite{Braden:1999aj,Mironov:1999vj}.

The genus-2 Riemann theta function with elliptic parameters $p$ and $w$ explicitly has the following form
\begin{align}
\Theta\left[\begin{array}{cc} a & b \\ \frac{1}{2} & 0 \end{array}\right]&\left( \mathsf{q}  \ \ \ g \mathsf{p} \left| \begin{array}{cc} \tau_1 & g m \\ g m & \tau_2 \end{array} \right. \right) \cr
&= \sum_{k_1, k_2 \in\mathbb{Z}} p^{\frac{(k_1+a)^2 - a^2}{2}}\, w^{(k_2+b)^2 - b^2}\, t^{ 2 (k_1 + a)(k_2 + b) } \notag\\
&\qquad\times\big(-{ Q}^2\big)^{ k_1 + a }\, \big({ P}^2\big)^{ k_2 + b }\,,
\label{Genus2ThetaDef}
\end{align}
where we introduced the following exponential parameters
\begin{align}
q = e^{\pi i g}, \ \ \ t = e^{\pi i m g}, \ \ \ p = e^{2\pi i \tau_1}, \ \ \ w = e^{\pi i \tau_2}, \ \ \ { P} = e^{\pi i g \mathsf{p}}, \ \ \ { Q} = e^{\pi i \mathsf{q}}\,.
\end{align}
In this rank-one operator formula $m$ is the mass in the additive convention $t=e^{\pi i gm}$, so the mixed Fourier factor is $t^{2(k_1+a)(k_2+b)}$. In \eqref{eq:gN2curve} the mass was measured by $\log t$. Here $p$ and $w$ are the two elliptic parameters, and $\mathsf{q},\mathsf{p}$ denote the canonical coordinate and momentum. The factor $p^{-a^2/2} w^{-b^2}$ normalizes the theta function so that it is a series in non-negative powers of $p^{1/2}$ and $w$.

Using \eqref{Genus2ThetaDef} we can show that for $N=2$ the only DELL Hamiltonian can be written in terms of genus-2 theta functions 
\begin{equation}
\mathcal{ \widehat{ H}} = \widehat{ \mathcal{ O}}_0^{-1} \widehat{ \mathcal{ O}}_1\,,
\end{equation}
where
\begin{equation}
\widehat{\mathcal{O}}_0 = -i\,\Theta\left[\begin{array}{cc} \frac{1}{2} & 0 \\ \frac{1}{2} & 0 \end{array}\right] \left(\frac{\widehat{x}_2}{\widehat{x}_1}, \frac{\widehat{p}_2}{\widehat{p}_1}\right)\,,\quad
\widehat{\mathcal{O}}_1 = i\,\Theta\left[\begin{array}{cc} \frac{1}{2} & \frac{1}{2} \\ \frac{1}{2} & 0 \end{array}\right]\left(\frac{\widehat{x}_2}{\widehat{x}_1}, \frac{\widehat{p}_2}{\widehat{p}_1}\right)\big(\widehat{p}_1\widehat{p}_2\big)^{1/2}\,.
\end{equation}
Here $Q^2$ and $P^2$ in \eqref{Genus2ThetaDef} are replaced by $\widehat{x}_2/\widehat{x}_1$ and $\widehat{p}_2/\widehat{p}_1$, with functions of $\widehat{x}$ standing to the left, and $(-Q^2)^{k+a}:=e^{\pi i(k+a)}Q^{2(k+a)}$. Hence $\widehat{\mathcal{H}}=-\Theta_0^{-1}\Theta_1\,(\widehat{p}_1\widehat{p}_2)^{1/2}$, where $\Theta_0$ and $\Theta_1$ denote the two theta functions above.
\smallskip\\
This gives a genus-two theta formula for the two-particle Hamiltonian. For arbitrary $N$, Grekov and Zotov give a characteristic determinant for the DELL current \cite{Grekov:2020jpk}.

\vspace{.2cm}
In the limit $w\to 0$ for $N=2$ we get from \eqref{eq:CurrentOz}
\begin{equation}
\mathcal{O}(z)= \vartheta\left(\frac{x_1}{x_2}\Big|p\right)-z \left(\vartheta \left(\frac{tx_1}{x_2}\Big|p\right)p_1+\vartheta \left(\frac{x_1}{t x_2}\Big|p\right)p_2 \right)+ z^2  \vartheta \left(\frac{x_1}{x_2}\Big|p\right)p_1 p_2 \,,
\end{equation}
which yields the integrals of motion of the two-body eRS model
\begin{equation}
\mathcal{H}_1=\mathcal{O}_0^{-1}\mathcal{O}_1=-\frac{\vartheta \left(t\frac{x_1}{x_2}\Big|p\right)}{\vartheta\left(\frac{x_1}{x_2}\Big|p\right)}\,p_1-\frac{\vartheta \left(t\frac{x_2}{x_1}\Big|p\right)}{\vartheta\left(\frac{x_2}{x_1}\Big|p\right)}\,p_2 \,, \qquad \mathcal{H}_2=\mathcal{O}_0^{-1}\mathcal{O}_2=p_1 p_2\,,
\end{equation}
where we used $\vartheta(1/x|p)=-\vartheta(x|p)$.

\vspace{.2cm}
In the limit $p\to 0$ we get from \eqref{eq:ShiftedTheta} $\vartheta(z|0)=z^{-1/2}(1-z)$ and \eqref{eq:CurrentOz} reads
\begin{align}
\mathcal{O}(z)&= \Big(\frac{x_2}{x_1}\Big)^{\frac12} \sum_{n\in\mathbb{Z}} w^{n^2}\, t^{-n}\left(1-t^{2n}\frac{x_1}{x_2}\right)\left(\frac{p_1}{p_2}\right)^{n}\notag \\
&-z\Big(\frac{x_2}{x_1}\Big)^{\frac12}\sum_{n\in\mathbb{Z}} w^{n^2-n}\, t^{-\frac{2n-1}{2}}\left(1-t^{2n-1}\frac{x_1}{x_2}\right)\left(\frac{p_1}{p_2}\right)^{n}p_2\notag\\
&+z^2\Big(\frac{x_2}{x_1}\Big)^{\frac12}\sum_{n\in\mathbb{Z}} w^{n^2}\, t^{-n}\left(1-t^{2n}\frac{x_1}{x_2}\right)\left(\frac{ p_1}{ p_2}\right)^{n} p_1 p_2 + \dots
\,,
\label{eq:p0limitexp}
\end{align}
and can be summed as follows
\begin{align}
\mathcal{O}(z) \ = \ \Big(\frac{x_2}{x_1}\Big)^{\frac12}\theta\big(z t^{-\frac12} p_1\big|w\big)\, \theta\big(z t^{\frac12} p_2\big|w\big) - \Big(\frac{x_1}{x_2}\Big)^{\frac12} \theta\big( z t^{\frac12} p_1\big|w \big)\, \theta\big( z t^{-\frac12} p_2\big|w\big)\,,
\label{eq:p0limitdell}
\end{align}
where $\theta(\cdot|w)$ is the theta function of \eqref{eq:ShiftedTheta} without the prefactor $x^{-1/2}$.
This yields a twisted version of the dual eRS Hamiltonians after expanding in modes
\begin{align}
\mathcal{O}_0 &= \Big(\frac{x_2}{x_1}\Big)^{\frac12}\theta\big( - w t^{-1} p_1/p_2\big|w^2\big) - \Big(\frac{x_1}{x_2}\Big)^{\frac12} \theta\big( - w t\, p_1/p_2\big|w^2\big)\,,\notag \\
\mathcal{O}_1 &= -\Big[t^{\frac12}\Big(\frac{x_2}{x_1}\Big)^{\frac12}\theta\big( - t^{-1} p_1/p_2\big| w^2 \big) - t^{-\frac12}\Big(\frac{x_1}{x_2}\Big)^{\frac12} \theta\big( - t\, p_1/p_2\big| w^2\big)\Big]p_2\,,
\end{align}
and the other modes can be obtained by periodicity.

\subsection{Exact Spectrum at \texorpdfstring{$p=0$}{p=0}}\label{sec:p0exact}
Grekov and Zotov obtained the triangular action and product spectrum below in Section~4 of \cite{Grekov:2020jpk}. Their spectral parameter is $z\,t^{-(N-1)/2}$ in our centered Weyl-vector convention. The relation of the resulting eigenpolynomials to elliptic RS duality is studied in \cite{MMZ:eRS}. At $p=0$ the coordinate dependence of \eqref{eq:CurrentOz} becomes trigonometric, $\vartheta(y|0)=y^{-1/2}-y^{1/2}$, and the spectrum can be found exactly. Let $\Delta(\textbf{x})=\prod_{i<j}\vartheta(x_i/x_j|0)$, let $m_\lambda$ denote the monomial symmetric functions, and let $\rho_i=\frac{N+1}{2}-i$.
\begin{proposition}\label{prop:p0}
At $p=0$ the operators $\Delta(\textbf{x})^{-1}\widehat{\mathcal{O}}_a$ preserve the space of symmetric polynomials of each degree and are triangular with respect to the dominance order,
\begin{equation}
\Delta(\textbf{x})^{-1}\widehat{\mathcal{O}}_a\, m_\lambda=\varepsilon_a(\lambda)\, m_\lambda+\sum_{\mu<\lambda}c^{(a)}_{\lambda\mu}\, m_\mu\,,\quad \sum_{a\in\mathbb{Z}}\varepsilon_a(\lambda)\,z^a=\prod_{i=1}^N\theta\big(z\,q^{\lambda_i}t^{\rho_i}\big|w\big)\,.
\label{eq:p0triangular}
\end{equation}
For generic $q,t$ and formal $w$ (or where $\varepsilon_0$ is nonzero on the relevant degree), each $\widehat{\mathcal{H}}_a=\widehat{\mathcal{O}}_0^{-1}\widehat{\mathcal{O}}_a$ acts triangularly on symmetric polynomials, with eigenvalues $\varepsilon_a(\lambda)/\varepsilon_0(\lambda)$. If the Hamiltonians commute at $p=0$ (Conjecture~\ref{Th:ConjComm}), then, for generic parameters with simple spectrum in each degree, they have common eigenpolynomials with leading term $m_\lambda$, and the eigenvalue current reads
\begin{equation}
\lambda(z)\big|_{p=0}=\frac{\prod_{i=1}^N\theta\big(z\,q^{\lambda_i}t^{\rho_i}\big|w\big)}{\big[z^0\big]\prod_{i=1}^N\theta\big(z\,q^{\lambda_i}t^{\rho_i}\big|w\big)}\,.
\label{eq:p0spectrum}
\end{equation}
\end{proposition}
\begin{proof}[Sketch of proof]
The coefficient of $\widehat{p}_1^{n_1}\cdots\widehat{p}_N^{n_N}$ in $\Delta^{-1}\widehat{\mathcal{O}}_a$ has at most simple poles at $x_i=x_j$. Their residues cancel between $\vec n$ and its transposition $s_{ij}\vec n$, which carry the same weight $w^{\sum_k n_k(n_k-1)/2}$; this gives polynomiality and symmetry. Expanding the coefficients in the region $|x_1|\gg|x_2|\gg\dots\gg|x_N|$ shows that $m_\lambda$ is mapped to itself plus lower terms. The leading coefficient of a term is $\prod_{i<j}t^{(n_i-n_j)/2}=\prod_i t^{\rho_i n_i}$, and summing over $\vec n$ with the signs and weights of \eqref{eq:CurrentOz} gives the product of theta functions.
\end{proof}
We have checked \eqref{eq:p0triangular}, together with $[\widehat{\mathcal{H}}_a,\widehat{\mathcal{H}}_b]=0$ on the resulting finite-dimensional spaces, in exact arithmetic for $N=2,3$ and degrees up to four. For $N=2$ and $\lambda=(1,0)$ only $m_{1,0}=x_1+x_2$ appears, and
\begin{equation}
\begin{split}
\lambda_1\big|_{p=0}&=-t^{-1/2}\,\frac{\theta(-qt|w^2)}{\theta(-wqt|w^2)}\\
&=-t^{-1/2}-t^{1/2}q+w\,\frac{(qt+1)(q^2t^2+1)}{t^{3/2}q}+O(w^2)\,,
\end{split}
\label{eq:U2p0}
\end{equation}
in agreement with \eqref{eq:Zellexp22}. For $N=2$ the eigenpolynomials are the elliptic Macdonald polynomials of Section~\ref{Sec:EllipticMacPoly}: with $x_1=x$ and $x_2=x^{-1}$, the eigenfunction with leading term $m_{(j,0)}$ equals $\mathcal{P}_j(x;q,t,w)$, i.e. the elliptic parameter of Section~\ref{Sec:EllipticMacPoly} is $w$. We have checked this in exact arithmetic through order $w^4$ for $j\leq 5$.

There is also a direct argument for every $j\geq0$. Set $z=t^{-1/2}q^{-j}$ in \eqref{eq:p0limitdell} and write a homogeneous solution as $\sum_{k=0}^j c_k x_1^{j-k}x_2^k$, with $c_0=1$. Comparing adjacent coefficients and using $\theta(y^{-1}\mid w)=-y^{-1}\theta(y\mid w)$ gives
\begin{equation*}
\frac{c_k}{c_{k-1}}=
\frac{\theta(q^{j-k+1}\mid w)\,\theta(tq^{k-1}\mid w)}
     {\theta(tq^{j-k}\mid w)\,\theta(q^k\mid w)}
\qquad(1\leq k\leq j).
\end{equation*}
The recurrence terminates at $k=j+1$ and gives precisely the polynomial of Section~\ref{Sec:EllipticMacPoly}. To recover the full eigenvalue equation, take $0<|w|<1$ and generic $q,t$. On homogeneous degree $j$, quasiperiodicity makes each coefficient of the current applied to this polynomial a degree-two theta section in $z$. The product $\theta(zt^{1/2}q^j\mid w)\theta(zt^{-1/2}\mid w)$ has the same multiplier. Vanishing at $z=t^{-1/2}q^{-j}$ cuts the two-dimensional space of sections down to one dimension. Each coefficient is therefore proportional to that product. Taking the constant coefficient in $z$ fixes the normalization in \eqref{eq:p0spectrum} and gives the identity to all orders in $w$.

\subsection{Quiver Generalization. Spin-DELL}\label{sec:spinDELL}
The above construction suggests a straightforward generalization to $\widehat{A}$-type 6d quiver gauge theories. Given an affine quiver theory of $L$ $U(N)$ gauge groups we propose the following generating function
\begin{multline}
\widehat{ \mathcal{O^S}}(z) = \sum_{n \in {\mathbb Z}} \widehat{ \mathcal{O^S}}_n\, z^n \\
= \sum_{\textbf{n}_{1},\dots,\textbf{n}_{L} \in \mathbb{Z}^N} (-z)^{\sum_{a,i} n_{a,i}}\, w^{\sum_{a,i} \frac{n_{a,i}(n_{a,i} - 1)}{2}} \\
\times\prod_{a=1}^L\prod_{i < j} \vartheta\left( t^{n_{a,i} - n_{a,j}} \frac{\widehat{x}_{a,i}}{\widehat{x}_{a,j}} \Big\vert p\right) \prod_{a=1}^L\prod_{i=1}^N\widehat{p}_{a,i}^{\,n_{a,i}}\,,
\label{eq:CurrentOzSpin}
\end{multline}
where subscript $a$ enumerates gauge groups in the quiver, $\textbf{n}_{a}=\{n_{a,1},\allowbreak\dots,\allowbreak n_{a,N}\}$, and the mode number is $n=\sum_{a,i}n_{a,i}$. We can call this model \textit{spin-DELL} analogously to the spin-Calogero system. In the brane construction coordinates $x_{a,i}$ show the position of endpoints of Lagrangian branes corresponding to $a$th gauge group. As written, \eqref{eq:CurrentOzSpin} contains no factors that couple different nodes, so it factorizes, $\widehat{\mathcal{O}^S}(z)=\prod_{a=1}^L\widehat{\mathcal{O}}^{(a)}(z)$, into mutually commuting copies of \eqref{eq:CurrentOz}. This describes $L$ decoupled DELL systems. Commutativity of the normalized product modes follows if the normalized Hamiltonians at each node commute. We checked $L=2$, $N=2$ through order $w^2$. A genuine spin-DELL model requires additional factors coupling $x_{a,i}$ to $x_{a+1,j}$, reflecting the bifundamental matter; we leave their construction for future work.

\section{Defect Partition Function and Spectrum}\label{sec:States}
Our task is to compute instanton partition functions of 6d $(1,0)^*$ theories with monodromy defects and relate them to the eigenfunctions of elliptic integrable systems. In \cite{Bullimore:2015fr} it was shown how to do this computation for 5d $\CN=1^*$ theory. The 6d $(1,0)^*$ theories are natural generalizations of
the corresponding 5d $\CN=1^*$ theories. In particular, the 6d theories with the same matter content as their 5d cousins are anomaly free.

\subsection{Instanton Partition Function}
Let us briefly review the essential ingredients of the instanton calculus \cite{Nekrasov:2002qd,Nekrasov:2003rj} with ramification \cite{Alday:2010vg,Nawata:2014nca,Bullimore:2015fr,Nekrasov:2017rqy}.
What follows represents a short summary of Section 4 of \cite{Bullimore:2015fr} albeit modified for the 6d theories on $\mathbb{R}^4\times T^2$ from the 5d theories on $\mathbb{R}^4\times S^1$.

For the ADHM description of $\mathcal M_{k,N}$ take $V=\mathbb C^k$, $W=\mathbb C^N$, $A,B\in\operatorname{End}(V)$, $I:W\to V$ and $J:V\to W$. The instanton moduli space is the stable quotient of $[A,B]+IJ=0$ by $GL(V)$.

The two rotations of $\mathbb C^2$ and the framing torus act on $\mathcal M_{k,N}$. Their maximal torus is $T=\mathbb C^\times_{q_1}\times\mathbb C^\times_{q_2}\times\mathbb T(GL(N;\mathbb C))$. Its fixed points are indexed by $N$-tuples of Young diagrams of total size $k$.

In the convention of Eqs.~(4.9)--(4.11) of \cite{Bullimore:2015fr}, the $T$-equivariant character of the universal bundle at the fixed point labeled by $\boldsymbol\lambda$ reads
\begin{equation}
\mathscr{U} = \mathscr{W}-(1-q_1^{-1})(1-q_2^{-1})\mathscr{V}\,,
\end{equation}
where
\begin{equation}
\mathscr{W}= \sum_{I=1}^N e^{a_I}\,,\qquad \mathscr{V}= \sum_{I=1}^N e^{a_I}\sum_{(r,c)\in\lambda_I} q_1^{1-r} q_2^{1-c}\,,
\end{equation}
where $q_1=e^{\epsilon_1},q_2=e^{\epsilon_2},e^{a_1},\dots,e^{a_N}$ are the equivariant parameters of $T$.
Using $\mathscr{U}$ we define the equivariant character of the tangent space to $\mathcal{M}_{k,N}$ at $\lambda$
\begin{equation}
\chi_1 = -(1-q_1)(1-q_2)\mathscr{V}\mathscr{V}^* + q_1 q_2 \mathscr{W} \mathscr{V}^* + \mathscr{W}^*\mathscr{V}\,,
\end{equation}
where the conjugation $^*$ corresponds to inverting the equivariant parameters. Due to the presence of the adjoint matter with mass $m$ we need to multiply the above expression by
\begin{equation}
\chi_2=(1 - Q_m q_1^{-1}q_2^{-1})\chi_1\,,
\label{eq:chi2def}
\end{equation}
where $Q_m=e^{m}$. The equivariant character can always be written as follows
\begin{equation}
\chi_2 = \sum_\alpha n_\alpha e^{\omega_\alpha}\,,
\end{equation}
from which we compute the 6d instanton partition function. The contribution to the topological sector with $k$ instantons reads
\begin{equation}
\CZ_k = \sum_{|\boldsymbol{\lambda}|=k}\prod_\alpha \vartheta(e^{\omega_\alpha}\mid w)^{-n_\alpha}\,,
\label{eq:NekPartFunContr}
\end{equation}
where the signed multiplicities are those of the virtual character $\chi_2$. We use the odd multiplicative theta function of \eqref{eq:ShiftedTheta} for every weight, both before and after orbifold projection. We identify the multiplicative elliptic parameter of the genus with the $w$ of the quantum current. In additive coordinates this choice gives $e^{2\pi i\tau_{\rm ell}}=w$, whereas the period $\tau_2$ in the classical theta matrix of Section~2 satisfies $e^{2\pi i\tau_2}=w^2$. The two-particle calculation below tests this choice against the spectrum. For the affine Laumon elliptic genus in multiplicative conventions see Section~8 of \cite{Awata:ELS}. Summing over the bulk instanton number gives (see also \cite{Kim:2015uq})
\begin{equation}
\CZ = \sum_k p^k \CZ_k\,,
\label{eq:6d4dpf}
\end{equation}
where $p$ is a dimensionless combination of the Yang-Mills coupling and the torus area.

\subsection{Ramification}
Now we need to upgrade the construction by introducing a defect \cite{Alday:2010vg,Bullimore:2015fr}.
The codimension-two defect wrapping $\mathbb{C}_{q_1}$ plane is described by specifying the eigenvalues of the $U(N)$ gauge field monodromy
\begin{equation}
\oint A^a = 2\pi m_a, \qquad a=1,\dots,N\,,
\end{equation}
where
\begin{equation}
m^a = (\underbrace{m_1,\dots,m_1}_{n_1},\underbrace{m_2,\dots,m_2}_{n_2},\dots,\underbrace{m_s,\dots,m_s}_{n_s})
\end{equation}
In the presence of such a defect the symmetry is broken to the Levi subgroup $U(n_1)\times\dots\times U(n_s)$ of $U(N)$. The full defect has $s=N$. The simple defect has Levi type $(N-1,1)$, and $s=1$ is the unramified case.

The $\mathbb Z_s$-fixed locus of the charge-$k$ ADHM space on the cover decomposes according to the dimensions $k_i$ of the color spaces in $V$,
\begin{equation}
\mathcal{M}_{k,N}^{\mathbb Z_s} =
 \bigsqcup_{k_1+\cdots+k_s=k}\mathcal{M}_{k_1,\dots,k_s,N}\,.
\end{equation}
We keep the mass invariant and project the virtual character \eqref{eq:chi2def} to the invariant weights as in Section~4.2 of \cite{Bullimore:2015fr},
\begin{equation}
\chi_{\rho,\boldsymbol{\lambda}}(a,\epsilon_1,\epsilon_2) = \frac{1}{s}\sum\limits_{r=1}^s\chi_2\left(a_{j,\alpha}-\frac{\epsilon_2+2\pi i r}{s}j,\epsilon_1,\frac{\epsilon_2+2\pi i r}{s}\right)=\sum\limits_{\alpha}n_\alpha e^{\omega_\alpha}\,,
\end{equation}
where $\rho$ specifies the Levi subgroup. Equivalently, $\chi_1=q_1q_2\chi_1^*$, so $\chi_2=\chi_1-Q_m\chi_1^*$. Taking invariants commutes with dualization. Thus each invariant tangent weight $y$ contributes $\vartheta(Q_m/y\mid w)/\vartheta(y\mid w)$.

Write $\lambda_{j,\alpha}$ for the diagrams of framing color $j$, with $\alpha=1,\ldots,n_j$. Each box in column $c$ has color $j+c-1$ modulo $s$, with representatives $1,\ldots,s$. Let $k_i=k_i(\boldsymbol\lambda)$ count the boxes of color $i$. In the chamber used below the sector fugacities, in color order, are
\begin{equation*}
 \frac{x_1}{x_2},\,\ldots,\,\frac{x_{s-1}}{x_s},\,p\,\frac{x_s}{x_1}.
\end{equation*}
Only the ratios enter, so we may take $x_1\cdots x_s=1$ for the relative positions. The closing sector has color $s$. The fractional-instanton sum is
\begin{equation}
\mathcal{Z}^{6d/4d}_{\text{inst}}
 =\sum_{\boldsymbol{\lambda}}\left(p\frac{x_s}{x_1}\right)^{k_s}
     \prod_{i=1}^{s-1}\left(\frac{x_i}{x_{i+1}}\right)^{k_i}
     \CZ_{\rho,\boldsymbol{\lambda}}\,,
\label{eq:pfdefect}
\end{equation}
where
\begin{equation}
\CZ_{\rho,\boldsymbol{\lambda}} = \prod_\alpha \vartheta(e^{\omega_\alpha}\mid w)^{-n_{\alpha}}\,.
\end{equation}
Adding one box of every color adds the regular representation and multiplies the sector weight by $p$. At $p=0$ only $k_s=0$ survives. For $s=1$ the sole fugacity is $p$ and the sum agrees term by term with \eqref{eq:6d4dpf}. At $s=2$ this is the $(z,p/z)$ convention of Eqs.~(4.25)--(4.27) of \cite{Bullimore:2015fr} with $z=x_1/x_2$.

\subsection{Eigenfunctions and the Difference Equation}
For the full defect take the Nekrasov--Shatashvili limit $q_2\to1$ after dividing by the bulk partition function. The quotient depends only on position ratios. A wavefunction in a specified chamber also has a leading monomial. With exponents $\lambda_i$ we write the proposed state, up to a scalar independent of $\mathbf x$, as
\begin{equation}
\mathscr{Z}(\textbf{x},\textbf{a},q_1,Q_m,p,w)
 =\left(\prod_{i=1}^N x_i^{\lambda_i}\right)
  \lim_{\epsilon_2\to0}\dfrac{\mathcal{Z}^{6d/4d}_{\text{inst}}}{\CZ}\,,
\label{eq:gaugetheoryeigenfunction}
\end{equation}
where the numerator and denominator are given by \eqref{eq:pfdefect} and \eqref{eq:6d4dpf}. The adjacent ratios $e^{a_i-a_{i+1}}=tq^{\lambda_{i+1}-\lambda_i}$ fix the differences of exponents. Their common shift specifies the center-of-mass character.

Conjecture~\ref{th:Conj2} predicts that $\mathscr{Z}$ satisfies the DELL difference equation
\begin{align}
\widehat{\mathcal{O}}(z) \mathscr{Z} = \lambda(z) \ \widehat{ \mathcal{O}}_{0} \mathscr{Z}\,,
\label{eigenproblemZ}
\end{align}
where the eigenvalue can be expanded in $z$-modes $\lambda(z)=\sum_{r}z^r\, \lambda_r(\textbf{a},\allowbreak q_1,\allowbreak Q_m,\allowbreak p,\allowbreak w)$, where each coefficient is a double series in the elliptic parameters
\begin{equation}
\lambda_r=\sum_{m,n} p^m w^n \lambda_{r,m,n} (\textbf{a},q_1,Q_m)\,,
\label{eq:lambdaexpansion}
\end{equation}
where $\lambda_{r,m,n} (\textbf{a},q_1,Q_m)$ are some rational functions of its arguments (see below and in \appref{sec:U3DEll}). The current $\lambda(z)$ possesses the same quasi-periodicity properties as the operator $\widehat{\mathcal{O}}(z)$ \eqref{eq:Oquasiperiod}.

\vspace{.5cm}

\subsection{Comments on Eigenvalues}\label{Sec:GuessEigen}
At the moment we do not have an independent localization computation of the eigenvalues $\lambda_r$ for the complete DELL system. However, we can find the terms in the series \eqref{eq:lambdaexpansion} order-by-order from the solution of the difference equation. Remarkably these terms are uniquely fixed by difference equation \eqref{eigenproblemZ} provided that we know the eigenfunction \eqref{eq:gaugetheoryeigenfunction}.

In the eRS limit $w\to 0$ the eigenvalues $\lambda_r$ are given by Proposition \ref{Conj1}; we have verified on a computer that the gauge-theory eigenfunctions satisfy this relation.

We plan to understand physical and geometrical nature of the difference equation \eqref{eigenproblemZ} in the near future. As a possible explanation the following argument may be useful. As we reviewed in the introduction, T-duality in Type IIB string theory rotates the brane web \figref{fig:toricgiag}. The left hand side of \eqref{eigenproblemZ} $\widehat{\mathcal{O}}_r \mathscr{Z} $ represents a formal shift operator in the defect Kahler parameters $\textbf{x}$. This corresponds to shifting the Lagrangian brane in \figref{fig:toricgiag} horizontally. After applying the T-duality the Lagrangian brane will end on the NS brane. Presumably the right hand side of \eqref{eigenproblemZ} $(\lambda_r \mathcal{O}_0 )\mathscr{Z}$ describes a result of the action of the shift operator in the vertical direction if we treat $(\lambda_r \mathcal{O}_0)$ as a whole object. This way the DELL difference equation will be a manifestation of the T-duality (p-q duality) similar to the 3d setup discussed in \cite{Gaiotto:2013bwa}, where the difference equation for the trigonometric RS model was equivalent to S-duality (3d mirror symmetry), which was the string realization of the p-q duality in three dimensions.

In this paper we only consider 6d theories with single unitary gauge group. One can also study quiver gauge theories. In type A there is a distinguished theory which exhibits p-q self-duality. Indeed, the toric diagram for the quiver theory with $M$ $U(M)$ gauge groups will be manifestly symmetric under the rotations by 90 degrees. 
This should be investigated further in more details. In particular, we expect that a coupled version of the spin-DELL Hamiltonians of Section~\ref{sec:spinDELL} will obey difference equations similar to \eqref{eigenproblemZ}.

\subsection{Truncation of $\mathscr{Z}$ state. Double Elliptic Macdonald Functions}
Since handling formal series in multiple variable is practically cumbersome, for most of the calculations we shall use truncated versions of the partition function \eqref{eq:gaugetheoryeigenfunction}.
We shall denote it by $\mathcal{P}_{{\textbf{j}}}(\textbf{x})$ and refer to it as the $U(N)$ DELL Macdonald function with elliptic parameters $p,w$; Macdonald parameters $q,t$; index ${\textbf{j}} = (j_1, \ldots, j_{N})$ and argument $\vec x = (x_1, \ldots, x_N)$.
Take $j_1\geq\cdots\geq j_N$ and expand in $x_i/x_{i+1}$ for $i<N$. The leading monomial in this chamber has exponents $\lambda_i=j_{N+1-i}$. We use the following specialization of the multiplicative parameters in the fixed-point character. The $e^{a_i}$ are specified up to a common factor.
\begin{equation*}
\begin{array}{c|c}
\text{Gauge theory} & \text{DELL parameters} \\
\hline
\text{Coulomb weights} & e^{a_i}\propto q^{-\lambda_i}t^{N-i} \\
\text{Sector fugacities} & x_i/x_{i+1}\ (i<N),\quad p\,x_N/x_1 \\
\text{Rotation and adjoint mass} & q_1=q,\quad Q_m=q/t \\
\text{Bulk instanton and elliptic genus parameters} & p,\quad e^{2\pi i\tau_{\rm ell}}=w
\end{array}
\end{equation*}

The adjacent ratios can be derived at $p=0$. Set all but one of the finite charges to zero. A fixed point of charge $k$ in color $i<N$ consists of a single column of height $k$ in diagram $i$. Substitution in the projected tangent character gives weights $q^d$ and $q^d e^{a_{i+1}-a_i}$ for $1\leq d\leq k$. Write $J=\lambda_{i+1}-\lambda_i$. At $e^{a_i-a_{i+1}}=tq^J$ the ratio of consecutive contributions is
\begin{equation}
 \frac{\theta(tq^{k-1}\mid w)\,\theta(q^{J-k+1}\mid w)}
      {\theta(q^k\mid w)\,\theta(tq^{J-k}\mid w)} .
 \label{eq:truncation-weights}
\end{equation}
Here odd inversion cancels the two signs, and the square-root factors cancel between numerator and denominator. The coefficient at $k=J+1$ vanishes. For $N=2$ these are all the fixed points at $p=0$. After multiplying by the chamber monomial we recover the entire elliptic Macdonald polynomial of Section~\ref{sec:p0exact} in every degree.

At arbitrary $N$, restrict the current to the same ray by setting the other adjacent position ratios to zero after division by the trigonometric Vandermonde. The only nonconstant position factor is $(1-t^{n_i-n_{i+1}}x_i/x_{i+1})/(1-x_i/x_{i+1})$. At $z=(q^{\lambda_i}t^{-\rho_i})^{-1}$ the eigenvalue in \eqref{eq:p0spectrum} vanishes. The spectator theta factors cancel and the two remaining factors give \eqref{eq:truncation-weights}. This checks each simple-root direction. Identifying the full series, including mixed charges at arbitrary rank, is part of Conjecture~\ref{th:Conj2}.

We have verified using computer algebra that the double elliptic Macdonald functions satisfy DELL difference equation
\begin{align}
\widehat{ \mathcal{O}}(z) \mathcal{ P}_{{\textbf{j}}}(\textbf{x}) = \lambda_{{\textbf{j}}}(z) \ \widehat{ \mathcal{O}}_{0} \mathcal{ P}_{{\textbf{j}}}(\textbf{x})\,,
\label{eigenproblemP}
\end{align}
where $\lambda_{{\textbf{j}}}(z)$ does not depend on $\textbf{x}$. The explicit spectral checks use the candidate polynomials in Section~\ref{sec:U2ex} and Appendix~\ref{sec:U3DEll}. Alternatively we can rewrite \eqref{eigenproblemP} as follows
\begin{align}
\left( \ \dfrac{d}{dz} \ \lambda^{-1}_{{\textbf{j}}}(z) \ \mathcal{L}(z) \ \right) \ \mathcal{ P}_{{\textbf{j}}}(\textbf{x}) = 0\,.
\end{align}

Note that the eigenvalue  $\lambda_{{\textbf{j}}}(z)$ is uniquely fixed by the difference equation \eqref{eigenproblemP}

\subsection{$U(2)$ Example}\label{sec:U2ex}
At $\textbf j=(1,0)$ take $(e^{a_1},e^{a_2})=(qt,1)$. The spectral factors in \eqref{eq:p0spectrum} are $(qt^{1/2},t^{-1/2})$. Their product is $q$, the eigenvalue of the simultaneous shift. At $p=0$ the two sector contributions in \eqref{eq:gaugetheoryeigenfunction} give $x_2(1+x_1/x_2)=x_1+x_2$.

The first term involving the closing sector already tests both elliptic parameters. The charge $(0,1)$ is a single box in diagram 2. After taking $q_2\to1$ its tangent weights are $q,q^2t$, giving the coefficient of $p x_2^2/x_1$ as
\begin{equation*}
 \frac{\vartheta(t\mid w)\,\vartheta(qt^2\mid w)}
      {\vartheta(q\mid w)\,\vartheta(q^2t\mid w)} .
\end{equation*}
The reflected monomial $p x_1^2/x_2$ comes from charge $(3,1)$. Of its four fixed points only $(\lambda_1,\lambda_2)=((1,1),(2))$, in row-length notation, has nonzero weight. Its projected virtual character coincides with that of $(0,1)$ even before the limit. Bulk division leaves both coefficients unchanged because the undressed coefficients of $(x_1/x_2)^{-1}$ and $(x_1/x_2)^2$ vanish at $p=0$. The charges $(1,1)$ and $(2,1)$ give equal coefficients through $pw$ after bulk division and have a finite limit. They contribute a scalar multiple of $x_1+x_2$ at this order.

For completeness, at $k_2=1$ the second diagram is either empty, a single box, or a row of length two. If it is empty, only the first row of diagram 1 can have length two or three, and any second row supplies a tangent weight $q^2e^{a_2-a_1}=Q_m$. If diagram 2 is nonempty, diagram 1 must be a column, and its third box supplies the same weight. In both cases $k_1\geq4$ forces a zero $\vartheta(1\mid w)$ in the numerator. This exhausts the sectors at order $p$.

A fixed-point product with a single $\vartheta(e^\omega\mid w^2)$ for each weight and $w$-independent fugacities would be even in $w$ and would miss the $pw$ coefficient. In the half-period convention one may equivalently split $\vartheta(y\mid w)$ into $\vartheta(y\mid w^2)\theta(wy\mid w^2)$ times a factor independent of $y$, by \eqref{eq:ShiftedTheta}. The latter factor cancels in the virtual character of rank zero. Expanding the theta quotient and normalizing the coefficient of $x_1+x_2$ to one gives
\begin{align}
\mathscr{Z}(\textbf{x})=\left(x_1+x_2\right)&+p\,\frac{q (t-1) \left(q t^2-1\right)}{t\,(q-1) \left(q^2t-1\right)}\left(\frac{x_1^2}{x_2}+\frac{x_2^2}{x_1}\right)\notag\\
&\times\left(1+w\,\frac{(q-t) (q t-1) (q t+1)^2}{q^2 t^2}+O(w^2)\right)+O(p^2)\,,
\label{eq:Zellexp2}
\end{align}
where the omitted terms are of higher order in $p$ and $w$. The corresponding eigenvalue reads
\begin{align}
&\lambda_1= -t^{-1/2}-t^{1/2}q-p\frac{(t-1)(qt+1)(q-t)^2}{t^{3/2}(q^2t-1)}\notag\\
&\quad + w\frac{(qt+1)(q^2t^2+1)}{t^{3/2}q}\notag\\
&\quad +p w\,\frac{(t-1)(qt+1)(q-t)^2}{t^{7/2}q^2(q^2t-1)}\notag\\
&\quad\times\big(q^4t^4-q^4t^3+q^3t^3+2q^3t+q^2t^3+2q^2t^2+q^2t\notag\\
&\qquad\quad+2qt^3+qt-t+1\big)+\dots\,.
\label{eq:Zellexp22}
\end{align}

In particular, if we put $p=w=0$ in \eqref{eq:Zellexp2} we get $\mathscr{Z}(\textbf{x})=x_1+x_2$, which is the first symmetric Macdonald polynomial.

\subsection{Perturbative Limit and Equivariant Elliptic Cohomology}
The double elliptic Macdonald functions \eqref{eq:gaugetheoryeigenfunction} are defined as formal Laurent series in $\textbf{x}$ variables. In the decoupling limit $p\to 0$ the coupled 6d/4d instanton/vortex system reduces to the 4d $\mathcal{N}=1$ defect theory on $\Reals^2\times T^2$.

At $p=0$ the instanton quotient in \eqref{eq:gaugetheoryeigenfunction} is a vortex series. In the flag-vertex description, suppressing the leading chamber monomial, the proposed series is
\begin{multline}
\left.\frac{\mathscr{Z}(\textbf{x},j_{\textbf{p}})}{\prod_i x_i^{\lambda_i}}\right|_{p=0} =
\sum\limits_{d_{i,j}\in C_{\textbf{p}}} \prod_{i=1}^{n-1} \left(\frac{q x_{i}}{t x_{i+1}}\right)^{d_i} \prod\limits_{j,k=1}^{i}\frac{\theta\left(q\frac{x_{i,j}}{x_{i,k}}\right)_{d_{i,j}-d_{i,k}}}{\theta\left(t\frac{x_{i,j}}{x_{i,k}}\right)_{d_{i,j}-d_{i,k}}}\\
\cdot\prod_{j=1}^{i}\prod_{k=1}^{i+1}\frac{\theta\left(t\frac{x_{i+1,k}}{x_{i,j}}\right)_{d_{i,j}-d_{i+1,k}}}{\theta\left(q\frac{x_{i+1,k}}{x_{i,j}}\right)_{d_{i,j}-d_{i+1,k}}}\,,
\label{eq:V1pdef}
\end{multline}
for a fixed flag $\textbf p$. The factor $(q/t)^{\sum_i d_i}$ results from converting the odd theta ratios to the unshifted $\theta$ in this formula. The factors within each node cancel in pairs. The factors between nodes $i$ and $i+1$ contribute an exponent $((i+1)d_i-i d_{i+1})/2$, which sums to $\sum_i d_i$ since $d_N=0$. We use the notation
\begin{equation}
\theta(x|w)_d=\prod\limits_{l=0}^{d-1}\theta(q^l x|w)\,.
\label{eq:thetaPoch}
\end{equation}
In \eqref{eq:V1pdef} chambers $C_{\textbf{p}}$ contain a collection  $\{d_{i,j}\}$ which satisfies the following property.
The collection of $d_{i,j} \geq 0$ must satisfy the following condition that for each $i = 1,\dots, n-2$ there should exist a subset in $\{d_{i+1,1}, . . . d_{i+1,i+1}\}$ of cardinality $i$, call it $\{d_{i+1,j_1},\dots, d_{i+1,j_i} \}$, such that $d_{i,k} \geq d_{i+1,j_k}$. Here $n=N$, $d_i=\sum_{j=1}^i d_{i,j}$, and $x_{i,j}$ are the torus weights of the rank-$i$ tautological subspace at the fixed flag $\textbf{p}$. At the top rank $d_{N,k}=0$ and the weights are $e^{a_1},\dots,e^{a_N}$. For $d>0$ set $\theta(x\mid w)_{-d}=1/\theta(q^{-d}x\mid w)_d$.

It is expected that such elliptic hypergeometric functions describe equivariant \textit{elliptic cohomology} of the cotangent bundle to the complete flag variety $T^*\mathbb{F}l_N$. Note that the above expression differs from that of the K-theoretic vertex function (Corollary 2.4 of \cite{Koroteev:2018isw}, see also \cite{Koroteev:2017nab}) only by replacing q-factorial symbols with theta functions $\theta(x|w)$. Analogously to vertex functions in the equivariant K-theory \eqref{eq:V1pdef} can be rewritten as a contour integral. Some development in this direction was made in \cite{Etingof:aa,Etingof:ab}.

We expect \eqref{eq:V1pdef} to give eigenfunctions for the coordinate-trigonometric limit of the normalized DELL Hamiltonians. For $N=2$, interchanging the two positions gives the opposite chamber. The eigenfunction of \eqref{eq:DefDellHam} at $p=0$ of the form $x_1^{\lambda_1}x_2^{\lambda_2}\sum_{k\geq 0}c_k\,(x_2/x_1)^k$ has
\begin{equation*}
\frac{c_k}{c_{k-1}}=\frac{q}{t}\,\frac{\theta(tq^{k-1}|w)\,\theta(Aq^{k-1}|w)}{\theta(q^k|w)\,\theta(Aq^k/t|w)}\,,\qquad A=q^{\lambda_2-\lambda_1}\,,
\end{equation*}
so it is an elliptic hypergeometric series of the type \eqref{eq:V1pdef}, built from the elliptic Pochhammer symbols \eqref{eq:thetaPoch}.

Let us look at the rank one theory in more detail. The 4d Nekrasov partition function for the $\mathcal{N}=1$ $U(1)$ SQED with two flavors on $\Reals^2\times T^2$ is an elliptic hypergeometric
function of the form \eqref{eq:V1pdef} for $N=2$. In \cite{Nieri:2015yia} it was shown that
the so-called \textit{holomorphic block,} which differs from the above vortex function by a  perturbative factor, is annihilated by the following operator
\begin{equation}
B_1=x\frac{\theta(t^2 p_1a^2\vert w)}{\theta(p_1a^2\vert w)}-x^{-1}\frac{\theta(t^2 p_1\vert w)}{\theta(p_1\vert w)}\,.
\label{eq:H1ellchiralring}
\end{equation}
Comparison with \eqref{eq:p0limitdell} also requires the perturbative factor relating the vortex function to the block. The elliptic holomorphic blocks naturally arise in the study of representations of the elliptic Virasoro algebra \cite{Nieri:2015dts}. In \eqref{eq:H1ellchiralring} we use the multiplicative $\theta$ of \eqref{eq:ShiftedTheta}. Replacing it by $\vartheta$ multiplies both terms of $B_1$ by $t^{-1}$ and preserves the annihilation equation.

We can recognize in the classical limit of $B_1\mathscr{Z}=0$, where $B_1$ is as in \eqref{eq:H1ellchiralring}, the twisted chiral ring relation of the theory which is quantized as
\begin{equation}
\sigma \to p_1\,.
\end{equation}
According to the Nekrasov-Shatashvili conjecture the above twisted chiral ring is isomorphic to the equivariant elliptic cohomology of $T^*\mathbb{P}^1$. These results can be extended to cotangent bundles to complete flag varieties $T^*\mathbb{F}l_n$. At the level of localization formulae all results from the equivariant K-theory translate directly to the elliptic cohomology by replacing q-factorials $(x;q)$ with theta functions $\theta(x|w)$.

We expect that a rigorous analysis of the equivariant elliptic cohomology of Nakajima quiver varieties
along the lines of \cite{Aganagic:2016hc} will be conducted in the context of the elliptic integrable systems in the near future.

\section{Elliptic Macdonald polynomials}\label{Sec:EllipticMacPoly}
It is worth discussing separately properties of the elliptic Macdonald polynomials which arise as truncation of series \eqref{eq:V1pdef}. This section can be read independently from the rest of the paper.

The elliptic Macdonald polynomials form a distinguished basis in the space of symmetric polynomials that corresponds to: physically, flag-type 4d holomorphic blocks on ${\mathbb C} \times S^1 \times S^1$ for some special values of the twisted mass parameters, mathematically, elliptic cohomology of flag varieties (again, for some special values of the equivariant parameters). In this section we summarize, in the case of root type ${\mathfrak g} = A_1$, their main properties: orthogonality and integrability.

\subsection{Definition}
Given complex parameters $q$ and $t = q^{\beta}$, let
\begin{align}
{\mathcal P}_{j}\big(x\big) := {\mathcal P}_{j}\big(x; q,t,p\big) = \sum\limits_{\ell = 0}^{j} \ x^{j - 2\ell} \ \prod\limits_{i = 0}^{\ell-1} \frac{[j - i]}{[j - i + \beta - 1]} \frac{[i + \beta]}{[i + 1]}\,,
\end{align}
where the bracket denotes the elliptic number,
\begin{align}
[n] = \dfrac{\theta(q^n|p)}{\theta(q|p)},
\end{align}
where the theta function $\theta(x|p)$ is defined in \eqref{eq:ShiftedTheta}. In the DELL system the elliptic parameter $p$ of this section is the momentum parameter $w$: for $N=2$, with $x_1=x$ and $x_2=x^{-1}$, $\mathcal{P}_j(x;q,t,w)$ is the eigenfunction of the Hamiltonians \eqref{eq:DefDellHam} at $p=0$ (Section~\ref{sec:p0exact}).

\subsection{Orthogonality}
For each elliptic Macdonald polynomial $\mathcal{P}_j$ we seek a dual Laurent series $\mathcal{P}^\star_j$ for the constant-term pairing
\begin{align}
\dfrac{1}{2!} \ \oint\limits_{|x|=1} \ \dfrac{dx}{2\pi i x} \ \ \ {\mathcal P}_{j}\big(x\big) \ {\mathcal P}^{\star}_{j^{\prime}}\big(x\big) = \delta_{j, j^{\prime}}
\label{EllOrtho}
\end{align}
Write $c_{j,\ell}=[x^{j-2\ell}]\mathcal P_j(x)$ for the coefficients in Section~5.1, so $c_{j,\ell}=c_{j,j-\ell}$. In the coefficients $c_{j,\ell}$ and $b_{k,s}$ below the elliptic parameter of the brackets is $w$. The dual coefficients, including those of the vacuum, are
\begin{align}
 \mathcal P^\star_k(x)&=\sum_{s\geq0} b_{k,s}\bigl(x^{k+2s}+x^{-k-2s}\bigr),\qquad b_{k,0}=1,\notag\\
 b_{k,s}&=t^s[k+2s]\prod_{i=1}^{s-1}[k+i]\,
  \prod_{i=0}^{s-1}\frac{[i-\beta]}{[k+\beta+1+i][i+1]},
 \qquad s\geq1 .
 \label{eq:momentum-dual}
\end{align}
For $k>0$ this is the expression obtained by canceling $[k]$ between the numerator and denominator. The same expression is regular at $k=0$ for generic parameters, and its $s=0$ term then contributes $2$ to $\mathcal P_0^\star$.

To verify \eqref{EllOrtho}, the only overlaps that need computation have $j=k+2n$, $n\geq0$. Symmetry of the coefficients and the Weyl factor $1/2$ give the overlap as $\sum_{s=0}^n b_{k,s}c_{k+2n,n-s}$. For $n=0$ this equals one, also for the vacuum. For $n>0$, dividing by $c_{k+2n,n}$ and using $\theta(y^{-1}\mid w)=-y^{-1}\theta(y\mid w)$ gives the terminating elliptic sum
\[
 {}_8V_7(q^k\,;\,t^{-1},tq^{k+n},q^{-n}\,;\,q,w)=0.
\]
This is the specialization of the Frenkel--Turaev summation given in Eq.~(2.1) of \cite{WarnaarEllipticSums}. The product evaluation contains the factor $\prod_{i=0}^{n-1}\theta(q^{1-n+i}\mid w)$, whose last term vanishes. For $k=0$ one first cancels the common theta factor in each positive-$s$ term as in \eqref{eq:momentum-dual}, then takes $q^k\to1$ in this finite identity. This proves biorthogonality coefficientwise.

Equation \eqref{EllOrtho} provides an elliptic generalization of the standard Macdonald orthogonality. In the limit $w\to0$ we have
\begin{align}
{\mathcal P}_{j}\big(x; q,t,0\big) = P_j\big(x\big), \qquad {\mathcal P}^{\star}_{j}\big(x; q,t,0\big) = \Delta_{q,t}(x) \ \dfrac{P_j\big(x\big)}{||P_j||^2}
\end{align}
where $\Delta_{q,t}(x)$ is the Macdonald weight
\begin{equation}
\Delta_{q,t}(x)=\frac{(x^2;q)_\infty\,(x^{-2};q)_\infty}{(tx^2;q)_\infty\,(tx^{-2};q)_\infty}\,,
\end{equation}
and
\begin{align}
P_{j}\big(x\big) = \sum\limits_{\ell = 0}^{j} \ x^{j - 2\ell} \ \prod\limits_{i = 0}^{\ell-1} \frac{(1 - q^{j - i})}{(1 - q^{j - i + \beta - 1})} \frac{(1 - q^{i + \beta})}{(1 - q^{i + 1})}
\end{align}
are the usual symmetric Macdonald polynomials. The relation (\ref{EllOrtho}) then reduces to
\begin{align}
\dfrac{1}{2!} \ \oint\limits_{|x|=1} \ \dfrac{dx}{2\pi i x} \ \ \ \Delta_{q,t}(x) \ P_{j}\big(x\big) \ P_{j^{\prime}}\big(x\big) = ||P_j||^2 \ \delta_{j, j^{\prime}}
\end{align}
which is the Macdonald orthogonality (see \cite{Aganagic:2011sg} for details and notations). One can see that the product of the measure with a Macdonald polynomial is deformed nicely if considered together.

\subsection{Difference Equations}
The following operator
\begin{equation}
{\widehat {\mathcal H}} =  {\widehat x}\,\dfrac{q}{t}\, \theta\left(tq^{\frac{j}{2}}{\widehat p}|p\right) \theta\left(q^{-\frac{j}{2}}{\widehat p}|p\right) -  {\widehat x}^{-1} \theta\left(q^{\frac{j}{2}}{\widehat p}|p\right) \theta\left(t^{-1}q^{-\frac{j}{2}}{\widehat p}|p\right)
\end{equation}
annihilates elliptic Macdonald polynomials:
\begin{equation}
{\widehat {\mathcal H}} \ {\mathcal P}_{j}\big(x\big) = 0,
\end{equation}
where the conjugate Heisenberg generators ${\widehat x}$ and ${\widehat p}$ satisfy ${\widehat p} {\widehat x} = q^{\frac{1}{2}} {\widehat x} {\widehat p}$, i.e. ${\widehat p}\,f(x)=f(q^{\frac12}x)$. Equivalently, one may use the operator
\begin{align}
{\widehat {\mathcal H}}'={\widehat x}\, \dfrac{q}{t}\dfrac{\theta\left(q^{-\frac{j}{2}}{\widehat p}|p\right)}{\theta\left(\frac{q}{t} q^{-\frac{j}{2}}{\widehat p}|p\right)}  - {\widehat x}^{-1}\, \dfrac{\theta\left(q^{\frac{j}{2}}{\widehat p}|p\right) }{\theta\left(\frac{t}{q} q^{\frac{j}{2}}{\widehat p}|p \right)}\,,
\end{align}
which is equivalent to the dual elliptic Ruijsenaars operator $B_1$ \eqref{eq:H1ellchiralring}.

For the orthogonal functions, one instead has
\begin{equation}
{\widehat {\mathcal H}}^{\star} \ {\mathcal P}^{\star}_{j}\big(x\big) = 0,
\end{equation}
where
\begin{equation}
{\widehat {\mathcal H}}^{\star} = \left[{\widehat x}\,t  \,\theta\left(q^{\frac{j}{2}}{\widehat p}|p\right) \theta\left(q^{-\frac{j}{2}} t^{-1} {\widehat p}|p \right) - {\widehat x}^{-1} \,\theta\left(q^{-\frac{j}{2}}{\widehat p}|p\right) \theta\left(q^{\frac{j}{2}} t {\widehat p}|p\right) \right]\dfrac{1}{\theta\left( {\widehat p}^2|p\right)} \,.
\end{equation}

\subsection{The \texorpdfstring{$TST$}{TST} integral identity}\label{sec:TST}
Consider the ordinary $A_1$ Macdonald polynomials $P_j$ of Section~5.2 in the relative coordinate $(x_1,x_2)=(x,x^{-1})$. For $0<q,t<1$, the Gaussian in this coordinate is
\begin{equation}
\theta(-q^{1/4}x\mid q^{1/2})=\sum_{n\in\mathbb Z}q^{n^2/4}x^n
\label{eq:TSTGaussian}
\end{equation}
in the theta convention of \eqref{eq:ShiftedTheta}. The rank-one Cherednik--Macdonald--Mehta identity \cite{CherednikCMM,EtingofKirillovCMM} then takes the form
\begin{equation}
\frac12\oint_{|x|=1}\frac{dx}{2\pi i x}\,
  \Delta_{q,t}(x)\,\theta(-q^{1/4}x\mid q^{1/2})\,P_j(x)P_{j'}(x)
 =\frac{(t;q)_\infty}{(t^2;q)_\infty}\,(TST)_{j,j'}\,,
\label{eq:TSTpairing}
\end{equation}
for $j,j'\geq0$. In the convention $S_{0,0}=T_{0,0}=1$, the entries are
\begin{align}
T_{j,j'}&=\delta_{j,j'}\,q^{j^2/4}t^{j/2}\,, &
S_{j,j'}&=P_j(t^{1/2})\,P_{j'}(t^{1/2}q^{j/2})\,.
\label{eq:TSTentries}
\end{align}
The prefactor in \eqref{eq:TSTpairing} is the Gaussian integral at $j=j'=0$. Dividing every entry by that vacuum integral gives the matrix elements $T_{j,j}S_{j,j'}T_{j',j'}$ in the monic Macdonald basis used here \cite{Aganagic:2011sg}.

To see the power of $q$ in these formulas, first take $t=q^\beta$ with $\beta$ a positive integer. Apply the two-variable Gaussian identity in Eq.~(3) of \cite{Shakirov:2026cmm} to homogeneous polynomials of degrees $j,j'$. The half-sum and half-difference of the two logarithmic coordinates have variance $(\log q)/2$ in its real Gaussian convention $q>1$. The center-of-mass integral contributes $q^{(j+j')^2/4}$. Removing it leaves $q^{(j^2+j'^2)/4}t^{(j+j')/2}$ in the matrix element and the moment $q^{n^2/4}$ on $x^n$ in the relative coordinate. For integer $\beta$ the weight is a finite Laurent polynomial. The moment identity therefore continues algebraically to $0<q<1$, where \eqref{eq:TSTGaussian} realizes these moments on the unit circle.

The absolute normalization follows from $P_1(x)=x+x^{-1}$. The first vacuum-normalized entries from \eqref{eq:TSTentries} are $(TST)_{0,1}=q^{1/4}(1+t)$ and $(TST)_{1,1}=(1+t)(1+qt)$. Replacing $t$ by $qt$ multiplies the weight by $(1-tx^2)(1-tx^{-2})=(1+t)^2-tP_1(x)^2$. Thus the ratio of the vacuum integral at $qt$ to its value at $t$ is $(1+t)(1-qt^2)$. At $t=q$ the weight is $2-x^2-x^{-2}$ and the vacuum integral is $1-q$. Iteration gives $\prod_{m=\beta}^{2\beta-1}(1-q^m)$ at $t=q^\beta$, precisely the prefactor in \eqref{eq:TSTpairing}. For fixed $0<q<1$, both sides of that equation are holomorphic in $|t|<1$. In the product $T_{j,j}S_{j,j'}T_{j',j'}$ the half-powers of $t$ combine to nonnegative integer powers. The specializations $t=q^\beta$ accumulate at zero, giving \eqref{eq:TSTpairing} throughout this disk by continuation.

For the momentum-elliptic pair one can keep the Gaussian \eqref{eq:TSTGaussian} and compute its matrix in the biorthogonal basis. Denote the resulting integral by $M_{jk}(w)$. At $t=q^\beta$ with $\beta$ a positive integer, the factor $[i-\beta]$ in \eqref{eq:momentum-dual} sets $b_{k,s}=0$ for $s>\beta$. For all degrees $j,k\geq0$ and all $|w|<1$ away from poles, this gives the exact formula
\begin{align}
 M_{jk}(w)&=\frac12\oint_{|x|=1}\frac{dx}{2\pi i x}\,
       \theta(-q^{1/4}x\mid q^{1/2})\,\mathcal P_j(x)\mathcal P^\star_k(x)\notag\\
 &=\sum_{s=0}^{\beta}b_{k,s}
          \sum_{\ell=0}^j c_{j,\ell}\,q^{(j-2\ell+k+2s)^2/4}.
 \label{eq:momentum-gaussian}
\end{align}
Indeed the Gaussian moment on $x^m$ is $q^{m^2/4}$. The two powers $x^{\pm(k+2s)}$ give equal sums by $\ell\leftrightarrow j-\ell$, including the doubled constant for $k=s=0$. At $w=0$, \eqref{eq:TSTpairing} gives
\[
 M_{jk}(0)=\frac{(t;q)_\infty}{(t^2;q)_\infty}\,
            \frac{(TST)_{j,k}}{\|P_k\|^2}.
\]
For $0<q,t<1$ with generic $t$, the same expression with $s\geq0$ defines the matrix as a formal series in $w$. At each fixed order, expanding the theta products gives at most exponential growth in $s$. Each logarithmic coefficient is a finite sum of powers of the arguments and their inverses, and $b_{k,s}(0)=O(t^s)$. The factor $q^{s^2+O(s)}$ makes each coefficient sum absolutely convergent.

Even within a single column the elliptic correction depends on the row. For example, at $q=1/2$, $t=q^2$, \eqref{eq:momentum-gaussian} gives
\[
 q^{1/4}\frac{M_{11}(w)}{M_{01}(w)}=\frac98-\frac{27}{8}w+O(w^2).
\]
The ordinary evaluation gives $1+qt=9/8$, and any common normalization of the column cancels from this ratio. 
At this specialization there is also a two-column test for any common multiplicative kernel. Let $L_k$ be the linear functional on symmetric Laurent polynomials specified by $L_k(\mathcal P_j)=(TST)_{j,k}$ with the ordinary entries of \eqref{eq:TSTentries}. If one kernel realized these targets up to invertible column factors, evaluating it on $F\mathcal P^\star_1\mathcal P^\star_2$ in the two orders would make $L_1(F\mathcal P^\star_2)$ and $L_2(F\mathcal P^\star_1)$ proportional as functions of $F$. For the rows $F=1$ and $F=x+x^{-1}$, expanding in the monic basis and dividing the first row by $q^{1/4}$ gives
\[
 \begin{pmatrix}85+66w&32+10w\\77+45w&48+13w\end{pmatrix}
 \pmod{101,w^2}.
\]
The entries are rational to this order before reduction. The determinant is $43w$ in the indicated ring, contradicting proportionality. Equation~\eqref{eq:momentum-gaussian} supplies the mixed matrix elements directly. Whether they admit a factorization in terms of independently defined elliptic modular matrices and spectral evaluations remains open. The elliptic Gaussian identity in \cite{Shakirov:2026cmm} uses nonstationary Shiraishi functions and is verified there to first order in their elliptic parameter.

\appendix
\section{Spectrum of the 3-body DELL Model}\label{sec:U3DEll}
For $\textbf j=(1,0,0)$ the proposed $U(3)$ eigenfunction has the following expansion. The closing-color one-box term in \eqref{eq:pfdefect} gives an independent check. The chamber monomial is $x_3$ and $e^{a_1-a_3}=qt^2$, so the coefficient of $p x_3^2/x_1$ in the NS limit is
\begin{equation*}
 \frac{\vartheta(t\mid w)\,\vartheta(qt^3\mid w)}
      {\vartheta(q\mid w)\,\vartheta(q^2t^2\mid w)} .
\end{equation*}
Its constant and linear terms in $w$ agree with the coefficients of this monomial below. The elliptic parameters $p,w$ are in bold.
\vspace{0.2cm}

$
\mathscr{Z}(\textbf{x}) =x_{1}+x_{2}+x_{3}
$
\begin{center}
$+
\textbf{p}\dfrac{q (-1+t)}{(q-1) (q t-1)^3 (q t+1)^2 t\, x_1x_2x_3}\cdot
 (q^4 t^6 x_{1}^3 x_{2}+q^4 t^6 x_{1}^3 x_{3}+2 q^4 t^6 x_{1}^2 x_{2}^2+3 q^4 t^6 x_{1}^2 x_{2} x_{3}+2 q^4 t^6 x_{1}^2 x_{3}^2+q^4 t^6 x_{1} x_{2}^3+3 q^4 t^6 x_{1} x_{2}^2 x_{3}+3 q^4 t^6 x_{1} x_{2} x_{3}^2+q^4 t^6 x_{1} x_{3}^3+q^4 t^6 x_{2}^3 x_{3}+2 q^4 t^6 x_{2}^2 x_{3}^2+q^4 t^6 x_{2} x_{3}^3-2 q^4 t^5 x_{1}^2 x_{2} x_{3}-2 q^4 t^5 x_{1} x_{2}^2 x_{3}-2 q^4 t^5 x_{1} x_{2} x_{3}^2+2 q^3 t^6 x_{1}^2 x_{2} x_{3}+2 q^3 t^6 x_{1} x_{2}^2 x_{3}+2 q^3 t^6 x_{1} x_{2} x_{3}^2-q^4 t^4 x_{1}^2 x_{2}^2-2 q^4 t^4 x_{1}^2 x_{2} x_{3}-q^4 t^4 x_{1}^2 x_{3}^2-2 q^4 t^4 x_{1} x_{2}^2 x_{3}-2 q^4 t^4 x_{1} x_{2} x_{3}^2-q^4 t^4 x_{2}^2 x_{3}^2-q^3 t^5 x_{1}^3 x_{2}-q^3 t^5 x_{1}^3 x_{3}-q^3 t^5 x_{1}^2 x_{2}^2-q^3 t^5 x_{1}^2 x_{2} x_{3}-q^3 t^5 x_{1}^2 x_{3}^2-q^3 t^5 x_{1} x_{2}^3-q^3 t^5 x_{1} x_{2}^2 x_{3}-q^3 t^5 x_{1} x_{2} x_{3}^2-q^3 t^5 x_{1} x_{3}^3-q^3 t^5 x_{2}^3 x_{3}-q^3 t^5 x_{2}^2 x_{3}^2-q^3 t^5 x_{2} x_{3}^3-q^4 t^3 x_{1}^2 x_{2}^2-q^4 t^3 x_{1}^2 x_{3}^2-q^4 t^3 x_{2}^2 x_{3}^2+2 q^3 t^4 x_{1}^2 x_{2}^2+2 q^3 t^4 x_{1}^2 x_{3}^2+2 q^3 t^4 x_{2}^2 x_{3}^2-q^2 t^5 x_{1}^2 x_{2}^2-q^2 t^5 x_{1}^2 x_{3}^2-q^2 t^5 x_{2}^2 x_{3}^2+q^4 t^2 x_{1}^2 x_{2} x_{3}+q^4 t^2 x_{1} x_{2}^2 x_{3}+q^4 t^2 x_{1} x_{2} x_{3}^2-q^3 t^3 x_{1}^3 x_{2}-q^3 t^3 x_{1}^3 x_{3}-q^3 t^3 x_{1}^2 x_{2}^2-5 q^3 t^3 x_{1}^2 x_{2} x_{3}-q^3 t^3 x_{1}^2 x_{3}^2-q^3 t^3 x_{1} x_{2}^3-5 q^3 t^3 x_{1} x_{2}^2 x_{3}-5 q^3 t^3 x_{1} x_{2} x_{3}^2-q^3 t^3 x_{1} x_{3}^3-q^3 t^3 x_{2}^3 x_{3}-q^3 t^3 x_{2}^2 x_{3}^2-q^3 t^3 x_{2} x_{3}^3-q^2 t^4 x_{1}^3 x_{2}-q^2 t^4 x_{1}^3 x_{3}-3 q^2 t^4 x_{1}^2 x_{2}^2-3 q^2 t^4 x_{1}^2 x_{3}^2-q^2 t^4 x_{1} x_{2}^3-q^2 t^4 x_{1} x_{3}^3-q^2 t^4 x_{2}^3 x_{3}-3 q^2 t^4 x_{2}^2 x_{3}^2-q^2 t^4 x_{2} x_{3}^3-2 q t^5 x_{1}^2 x_{2} x_{3}-2 q t^5 x_{1} x_{2}^2 x_{3}-2 q t^5 x_{1} x_{2} x_{3}^2+2 q^3 t^2 x_{1}^2 x_{2} x_{3}+2 q^3 t^2 x_{1} x_{2}^2 x_{3}+2 q^3 t^2 x_{1} x_{2} x_{3}^2-2 q t^4 x_{1}^2 x_{2} x_{3}-2 q t^4 x_{1} x_{2}^2 x_{3}-2 q t^4 x_{1} x_{2} x_{3}^2+2 q^3 t x_{1}^2 x_{2} x_{3}+2 q^3 t x_{1} x_{2}^2 x_{3}+2 q^3 t x_{1} x_{2} x_{3}^2+q^2 t^2 x_{1}^3 x_{2}+q^2 t^2 x_{1}^3 x_{3}+3 q^2 t^2 x_{1}^2 x_{2}^2+3 q^2 t^2 x_{1}^2 x_{3}^2+q^2 t^2 x_{1} x_{2}^3+q^2 t^2 x_{1} x_{3}^3+q^2 t^2 x_{2}^3 x_{3}+3 q^2 t^2 x_{2}^2 x_{3}^2+q^2 t^2 x_{2} x_{3}^3+q t^3 x_{1}^3 x_{2}+q t^3 x_{1}^3 x_{3}+q t^3 x_{1}^2 x_{2}^2+5 q t^3 x_{1}^2 x_{2} x_{3}+q t^3 x_{1}^2 x_{3}^2+q t^3 x_{1} x_{2}^3+5 q t^3 x_{1} x_{2}^2 x_{3}+5 q t^3 x_{1} x_{2} x_{3}^2+q t^3 x_{1} x_{3}^3+q t^3 x_{2}^3 x_{3}+q t^3 x_{2}^2 x_{3}^2+q t^3 x_{2} x_{3}^3-t^4 x_{1}^2 x_{2} x_{3}-t^4 x_{1} x_{2}^2 x_{3}-t^4 x_{1} x_{2} x_{3}^2+q^2 t x_{1}^2 x_{2}^2+q^2 t x_{1}^2 x_{3}^2+q^2 t x_{2}^2 x_{3}^2-2 q t^2 x_{1}^2 x_{2}^2-2 q t^2 x_{1}^2 x_{3}^2-2 q t^2 x_{2}^2 x_{3}^2+t^3 x_{1}^2 x_{2}^2+t^3 x_{1}^2 x_{3}^2+t^3 x_{2}^2 x_{3}^2+q t x_{1}^3 x_{2}+q t x_{1}^3 x_{3}+q t x_{1}^2 x_{2}^2+q t x_{1}^2 x_{2} x_{3}+q t x_{1}^2 x_{3}^2+q t x_{1} x_{2}^3+q t x_{1} x_{2}^2 x_{3}+q t x_{1} x_{2} x_{3}^2+q t x_{1} x_{3}^3+q t x_{2}^3 x_{3}+q t x_{2}^2 x_{3}^2+q t x_{2} x_{3}^3+t^2 x_{1}^2 x_{2}^2+2 t^2 x_{1}^2 x_{2} x_{3}+t^2 x_{1}^2 x_{3}^2+2 t^2 x_{1} x_{2}^2 x_{3}+2 t^2 x_{1} x_{2} x_{3}^2+t^2 x_{2}^2 x_{3}^2-2 q x_{1}^2 x_{2} x_{3}-2 q x_{1} x_{2}^2 x_{3}-2 q x_{1} x_{2} x_{3}^2+2 t x_{1}^2 x_{2} x_{3}+2 t x_{1} x_{2}^2 x_{3}+2 t x_{1} x_{2} x_{3}^2-x_{1}^3 x_{2}-x_{1}^3 x_{3}-2 x_{1}^2 x_{2}^2-3 x_{1}^2 x_{2} x_{3}-2 x_{1}^2 x_{3}^2-x_{1} x_{2}^3-3 x_{1} x_{2}^2 x_{3}-3 x_{1} x_{2} x_{3}^2-x_{1} x_{3}^3-x_{2}^3 x_{3}-2 x_{2}^2 x_{3}^2-x_{2} x_{3}^3)
$
\vspace{0.1cm}

$
+\textbf{w p}\dfrac{ (-1+t) (t^3 q-1) (-t+q)}{q (q-1) (q t-1)^3 (q t+1)^2 t^4\, x_1x_2x_3}\cdot (q^6 t^8 x_{1}^3 x_{2}+q^6 t^8 x_{1}^3 x_{3}+q^6 t^8 x_{1}^2 x_{2}^2+2 q^6 t^8 x_{1}^2 x_{2} x_{3}+q^6 t^8 x_{1}^2 x_{3}^2+q^6 t^8 x_{1} x_{2}^3+2 q^6 t^8 x_{1} x_{2}^2 x_{3}+2 q^6 t^8 x_{1} x_{2} x_{3}^2+q^6 t^8 x_{1} x_{3}^3+q^6 t^8 x_{2}^3 x_{3}+q^6 t^8 x_{2}^2 x_{3}^2+q^6 t^8 x_{2} x_{3}^3-q^6 t^7 x_{1}^2 x_{2}^2-2 q^6 t^7 x_{1}^2 x_{2} x_{3}-q^6 t^7 x_{1}^2 x_{3}^2-2 q^6 t^7 x_{1} x_{2}^2 x_{3}-2 q^6 t^7 x_{1} x_{2} x_{3}^2-q^6 t^7 x_{2}^2 x_{3}^2+q^5 t^8 x_{1}^2 x_{2}^2+2 q^5 t^8 x_{1}^2 x_{2} x_{3}+q^5 t^8 x_{1}^2 x_{3}^2+2 q^5 t^8 x_{1} x_{2}^2 x_{3}+2 q^5 t^8 x_{1} x_{2} x_{3}^2+q^5 t^8 x_{2}^2 x_{3}^2-q^6 t^6 x_{1}^2 x_{2}^2-q^6 t^6 x_{1}^2 x_{2} x_{3}-q^6 t^6 x_{1}^2 x_{3}^2-q^6 t^6 x_{1} x_{2}^2 x_{3}-q^6 t^6 x_{1} x_{2} x_{3}^2-q^6 t^6 x_{2}^2 x_{3}^2-q^5 t^7 x_{1}^3 x_{2}-q^5 t^7 x_{1}^3 x_{3}+q^5 t^7 x_{1}^2 x_{2}^2-q^5 t^7 x_{1}^2 x_{2} x_{3}+q^5 t^7 x_{1}^2 x_{3}^2-q^5 t^7 x_{1} x_{2}^3-q^5 t^7 x_{1} x_{2}^2 x_{3}-q^5 t^7 x_{1} x_{2} x_{3}^2-q^5 t^7 x_{1} x_{3}^3-q^5 t^7 x_{2}^3 x_{3}+q^5 t^7 x_{2}^2 x_{3}^2-q^5 t^7 x_{2} x_{3}^3+q^4 t^8 x_{1}^2 x_{2} x_{3}+q^4 t^8 x_{1} x_{2}^2 x_{3}+q^4 t^8 x_{1} x_{2} x_{3}^2+q^6 t^5 x_{1}^2 x_{2}^2+2 q^6 t^5 x_{1}^2 x_{2} x_{3}+q^6 t^5 x_{1}^2 x_{3}^2+2 q^6 t^5 x_{1} x_{2}^2 x_{3}+2 q^6 t^5 x_{1} x_{2} x_{3}^2+q^6 t^5 x_{2}^2 x_{3}^2+q^5 t^6 x_{1}^3 x_{2}+q^5 t^6 x_{1}^3 x_{3}+3 q^5 t^6 x_{1}^2 x_{2}^2+q^5 t^6 x_{1}^2 x_{2} x_{3}+3 q^5 t^6 x_{1}^2 x_{3}^2+q^5 t^6 x_{1} x_{2}^3+q^5 t^6 x_{1} x_{2}^2 x_{3}+q^5 t^6 x_{1} x_{2} x_{3}^2+q^5 t^6 x_{1} x_{3}^3+q^5 t^6 x_{2}^3 x_{3}+3 q^5 t^6 x_{2}^2 x_{3}^2+q^5 t^6 x_{2} x_{3}^3-2 q^4 t^7 x_{1}^2 x_{2}^2-2 q^4 t^7 x_{1}^2 x_{3}^2-2 q^4 t^7 x_{2}^2 x_{3}^2-q^6 t^4 x_{1}^2 x_{2} x_{3}-q^6 t^4 x_{1} x_{2}^2 x_{3}-q^6 t^4 x_{1} x_{2} x_{3}^2-3 q^5 t^5 x_{1}^2 x_{2}^2-4 q^5 t^5 x_{1}^2 x_{2} x_{3}-3 q^5 t^5 x_{1}^2 x_{3}^2-4 q^5 t^5 x_{1} x_{2}^2 x_{3}-4 q^5 t^5 x_{1} x_{2} x_{3}^2-3 q^5 t^5 x_{2}^2 x_{3}^2-q^4 t^6 x_{1}^3 x_{2}-q^4 t^6 x_{1}^3 x_{3}+q^4 t^6 x_{1}^2 x_{2}^2+4 q^4 t^6 x_{1}^2 x_{2} x_{3}+q^4 t^6 x_{1}^2 x_{3}^2-q^4 t^6 x_{1} x_{2}^3+4 q^4 t^6 x_{1} x_{2}^2 x_{3}+4 q^4 t^6 x_{1} x_{2} x_{3}^2-q^4 t^6 x_{1} x_{3}^3-q^4 t^6 x_{2}^3 x_{3}+q^4 t^6 x_{2}^2 x_{3}^2-q^4 t^6 x_{2} x_{3}^3-2 q^3 t^7 x_{1}^2 x_{2} x_{3}-2 q^3 t^7 x_{1} x_{2}^2 x_{3}-2 q^3 t^7 x_{1} x_{2} x_{3}^2-2 q^5 t^4 x_{1}^2 x_{2}^2+q^5 t^4 x_{1}^2 x_{2} x_{3}-2 q^5 t^4 x_{1}^2 x_{3}^2+q^5 t^4 x_{1} x_{2}^2 x_{3}+q^5 t^4 x_{1} x_{2} x_{3}^2-2 q^5 t^4 x_{2}^2 x_{3}^2-2 q^4 t^5 x_{1}^3 x_{2}-2 q^4 t^5 x_{1}^3 x_{3}+q^4 t^5 x_{1}^2 x_{2}^2-7 q^4 t^5 x_{1}^2 x_{2} x_{3}+q^4 t^5 x_{1}^2 x_{3}^2-2 q^4 t^5 x_{1} x_{2}^3-7 q^4 t^5 x_{1} x_{2}^2 x_{3}-7 q^4 t^5 x_{1} x_{2} x_{3}^2-2 q^4 t^5 x_{1} x_{3}^3-2 q^4 t^5 x_{2}^3 x_{3}+q^4 t^5 x_{2}^2 x_{3}^2-2 q^4 t^5 x_{2} x_{3}^3-3 q^3 t^6 x_{1}^2 x_{2}^2-3 q^3 t^6 x_{1}^2 x_{3}^2-3 q^3 t^6 x_{2}^2 x_{3}^2+q^5 t^3 x_{1}^2 x_{2} x_{3}+q^5 t^3 x_{1} x_{2}^2 x_{3}+q^5 t^3 x_{1} x_{2} x_{3}^2+3 q^4 t^4 x_{1}^2 x_{2}^2-q^4 t^4 x_{1}^2 x_{2} x_{3}+3 q^4 t^4 x_{1}^2 x_{3}^2-q^4 t^4 x_{1} x_{2}^2 x_{3}-q^4 t^4 x_{1} x_{2} x_{3}^2+3 q^4 t^4 x_{2}^2 x_{3}^2+q^3 t^5 x_{1}^3 x_{2}+q^3 t^5 x_{1}^3 x_{3}-4 q^3 t^5 x_{1}^2 x_{2}^2+4 q^3 t^5 x_{1}^2 x_{2} x_{3}-4 q^3 t^5 x_{1}^2 x_{3}^2+q^3 t^5 x_{1} x_{2}^3+4 q^3 t^5 x_{1} x_{2}^2 x_{3}+4 q^3 t^5 x_{1} x_{2} x_{3}^2+q^3 t^5 x_{1} x_{3}^3+q^3 t^5 x_{2}^3 x_{3}-4 q^3 t^5 x_{2}^2 x_{3}^2+q^3 t^5 x_{2} x_{3}^3-4 q^2 t^6 x_{1}^2 x_{2} x_{3}-4 q^2 t^6 x_{1} x_{2}^2 x_{3}-4 q^2 t^6 x_{1} x_{2} x_{3}^2-3 q^4 t^3 x_{1}^2 x_{2}^2-q^4 t^3 x_{1}^2 x_{2} x_{3}-3 q^4 t^3 x_{1}^2 x_{3}^2-q^4 t^3 x_{1} x_{2}^2 x_{3}-q^4 t^3 x_{1} x_{2} x_{3}^2-3 q^4 t^3 x_{2}^2 x_{3}^2+3 q^2 t^5 x_{1}^2 x_{2}^2+q^2 t^5 x_{1}^2 x_{2} x_{3}+3 q^2 t^5 x_{1}^2 x_{3}^2+q^2 t^5 x_{1} x_{2}^2 x_{3}+q^2 t^5 x_{1} x_{2} x_{3}^2+3 q^2 t^5 x_{2}^2 x_{3}^2+4 q^4 t^2 x_{1}^2 x_{2} x_{3}+4 q^4 t^2 x_{1} x_{2}^2 x_{3}+4 q^4 t^2 x_{1} x_{2} x_{3}^2-q^3 t^3 x_{1}^3 x_{2}-q^3 t^3 x_{1}^3 x_{3}+4 q^3 t^3 x_{1}^2 x_{2}^2-4 q^3 t^3 x_{1}^2 x_{2} x_{3}+4 q^3 t^3 x_{1}^2 x_{3}^2-q^3 t^3 x_{1} x_{2}^3-4 q^3 t^3 x_{1} x_{2}^2 x_{3}-4 q^3 t^3 x_{1} x_{2} x_{3}^2-q^3 t^3 x_{1} x_{3}^3-q^3 t^3 x_{2}^3 x_{3}+4 q^3 t^3 x_{2}^2 x_{3}^2-q^3 t^3 x_{2} x_{3}^3-3 q^2 t^4 x_{1}^2 x_{2}^2+q^2 t^4 x_{1}^2 x_{2} x_{3}-3 q^2 t^4 x_{1}^2 x_{3}^2+q^2 t^4 x_{1} x_{2}^2 x_{3}+q^2 t^4 x_{1} x_{2} x_{3}^2-3 q^2 t^4 x_{2}^2 x_{3}^2-q t^5 x_{1}^2 x_{2} x_{3}-q t^5 x_{1} x_{2}^2 x_{3}-q t^5 x_{1} x_{2} x_{3}^2+3 q^3 t^2 x_{1}^2 x_{2}^2+3 q^3 t^2 x_{1}^2 x_{3}^2+3 q^3 t^2 x_{2}^2 x_{3}^2+2 q^2 t^3 x_{1}^3 x_{2}+2 q^2 t^3 x_{1}^3 x_{3}-q^2 t^3 x_{1}^2 x_{2}^2+7 q^2 t^3 x_{1}^2 x_{2} x_{3}-q^2 t^3 x_{1}^2 x_{3}^2+2 q^2 t^3 x_{1} x_{2}^3+7 q^2 t^3 x_{1} x_{2}^2 x_{3}+7 q^2 t^3 x_{1} x_{2} x_{3}^2+2 q^2 t^3 x_{1} x_{3}^3+2 q^2 t^3 x_{2}^3 x_{3}-q^2 t^3 x_{2}^2 x_{3}^2+2 q^2 t^3 x_{2} x_{3}^3+2 q t^4 x_{1}^2 x_{2}^2-q t^4 x_{1}^2 x_{2} x_{3}+2 q t^4 x_{1}^2 x_{3}^2-q t^4 x_{1} x_{2}^2 x_{3}-q t^4 x_{1} x_{2} x_{3}^2+2 q t^4 x_{2}^2 x_{3}^2+2 q^3 t x_{1}^2 x_{2} x_{3}+2 q^3 t x_{1} x_{2}^2 x_{3}+2 q^3 t x_{1} x_{2} x_{3}^2+q^2 t^2 x_{1}^3 x_{2}+q^2 t^2 x_{1}^3 x_{3}-q^2 t^2 x_{1}^2 x_{2}^2-4 q^2 t^2 x_{1}^2 x_{2} x_{3}-q^2 t^2 x_{1}^2 x_{3}^2+q^2 t^2 x_{1} x_{2}^3-4 q^2 t^2 x_{1} x_{2}^2 x_{3}-4 q^2 t^2 x_{1} x_{2} x_{3}^2+q^2 t^2 x_{1} x_{3}^3+q^2 t^2 x_{2}^3 x_{3}-q^2 t^2 x_{2}^2 x_{3}^2+q^2 t^2 x_{2} x_{3}^3+3 q t^3 x_{1}^2 x_{2}^2+4 q t^3 x_{1}^2 x_{2} x_{3}+3 q t^3 x_{1}^2 x_{3}^2+4 q t^3 x_{1} x_{2}^2 x_{3}+4 q t^3 x_{1} x_{2} x_{3}^2+3 q t^3 x_{2}^2 x_{3}^2+t^4 x_{1}^2 x_{2} x_{3}+t^4 x_{1} x_{2}^2 x_{3}+t^4 x_{1} x_{2} x_{3}^2+2 q^2 t x_{1}^2 x_{2}^2+2 q^2 t x_{1}^2 x_{3}^2+2 q^2 t x_{2}^2 x_{3}^2-q t^2 x_{1}^3 x_{2}-q t^2 x_{1}^3 x_{3}-3 q t^2 x_{1}^2 x_{2}^2-q t^2 x_{1}^2 x_{2} x_{3}-3 q t^2 x_{1}^2 x_{3}^2-q t^2 x_{1} x_{2}^3-q t^2 x_{1} x_{2}^2 x_{3}-q t^2 x_{1} x_{2} x_{3}^2-q t^2 x_{1} x_{3}^3-q t^2 x_{2}^3 x_{3}-3 q t^2 x_{2}^2 x_{3}^2-q t^2 x_{2} x_{3}^3-t^3 x_{1}^2 x_{2}^2-2 t^3 x_{1}^2 x_{2} x_{3}-t^3 x_{1}^2 x_{3}^2-2 t^3 x_{1} x_{2}^2 x_{3}-2 t^3 x_{1} x_{2} x_{3}^2-t^3 x_{2}^2 x_{3}^2-q^2 x_{1}^2 x_{2} x_{3}-q^2 x_{1} x_{2}^2 x_{3}-q^2 x_{1} x_{2} x_{3}^2+q t x_{1}^3 x_{2}+q t x_{1}^3 x_{3}-q t x_{1}^2 x_{2}^2+q t x_{1}^2 x_{2} x_{3}-q t x_{1}^2 x_{3}^2+q t x_{1} x_{2}^3+q t x_{1} x_{2}^2 x_{3}+q t x_{1} x_{2} x_{3}^2+q t x_{1} x_{3}^3+q t x_{2}^3 x_{3}-q t x_{2}^2 x_{3}^2+q t x_{2} x_{3}^3+t^2 x_{1}^2 x_{2}^2+t^2 x_{1}^2 x_{2} x_{3}+t^2 x_{1}^2 x_{3}^2+t^2 x_{1} x_{2}^2 x_{3}+t^2 x_{1} x_{2} x_{3}^2+t^2 x_{2}^2 x_{3}^2-q x_{1}^2 x_{2}^2-2 q x_{1}^2 x_{2} x_{3}-q x_{1}^2 x_{3}^2-2 q x_{1} x_{2}^2 x_{3}-2 q x_{1} x_{2} x_{3}^2-q x_{2}^2 x_{3}^2+t x_{1}^2 x_{2}^2+2 t x_{1}^2 x_{2} x_{3}+t x_{1}^2 x_{3}^2+2 t x_{1} x_{2}^2 x_{3}+2 t x_{1} x_{2} x_{3}^2+t x_{2}^2 x_{3}^2-x_{1}^3 x_{2}-x_{1}^3 x_{3}-x_{1}^2 x_{2}^2-2 x_{1}^2 x_{2} x_{3}-x_{1}^2 x_{3}^2-x_{1} x_{2}^3-2 x_{1} x_{2}^2 x_{3}-2 x_{1} x_{2} x_{3}^2-x_{1} x_{3}^3-x_{2}^3 x_{3}-x_{2}^2 x_{3}^2-x_{2} x_{3}^3) + \ldots
$
\end{center}
Notice that the series purely in $w$ is absent because we took the smallest possible weight $\textbf{j}$.

The corresponding eigenvalues are

$
\lambda_1 = -q t-1-t^{-1}
$

\begin{center}
$
-\dfrac{\textbf{p}}{(q t+1) (q t-1)^2 t^2} (-1+t) (t+1) (-t+q) (t^3 q^3-q^2 t^4-t^3 q^2+t^3 q+q^2 t-t^2 q-q t+t^2+t-1)
$
\vspace{0.15cm}

$
+\dfrac{\textbf{w}}{t^3 q} (t^6 q^3+t^5 q^3+t^5 q^2+2 q^2 t^4+t^3 q^2+t^4 q+2 t^3 q+2 t^2 q+q t+t^2+t+1)
$
\vspace{0.1cm}

$
+\dfrac{\textbf{w p}}{q^2 (q t+1) (q t-1)^2 t^5} (-t+q) (-1+t) (t+1) (t^9 q^7-t^{10} q^6-t^8 q^7+2 t^8 q^6-q^5 t^9+2 t^7 q^6-3 t^8 q^5-t^9 q^4+t^7 q^5-t^8 q^4+2 t^5 q^6+2 t^6 q^5-4 t^7 q^4+2 t^5 q^5-5 t^6 q^4+2 t^4 q^5-t^6 q^3+3 t^7 q^2+t^3 q^5-3 t^5 q^3+3 t^6 q^2+2 t^3 q^4-4 t^4 q^3+2 t^5 q^2+q t^6+t^2 q^4-2 t^3 q^3+2 q^2 t^4+3 t^5 q-t^2 q^3-t^3 q^2+q^3 t-4 t^2 q^2+2 t^3 q-2 q^2 t+2 t^2 q-q t-t^2-q+t) + \ldots
$
\end{center}

and

$
\lambda_2 = q t+q+t^{-1}
$

\begin{center}
$
- \dfrac{\textbf{p}}{(q t+1) (q t-1)^2 t^2} (-1+t) (t+1) (-t+q) (t^4 q^3-t^3 q^3-t^2 q^3+t^3 q^2+t^2 q^2-t^3 q-q^2 t+q t+q-t)
$
\vspace{0.1cm}

$
- \dfrac{\textbf{w}}{t^3 q} (t^6 q^3+t^5 q^3+t^4 q^3+t^5 q^2+2 q^2 t^4+2 t^3 q^2+t^2 q^2+t^3 q+2 t^2 q+q t+t+1)
$
\vspace{0.1cm}

$
- \dfrac{\textbf{w p}}{q^2 (q t+1) (q t-1)^2 t^5} (-1+t) (t+1) (-t+q) (t^9 q^7-t^{10} q^6-t^8 q^7-t^9 q^6+2 t^8 q^6-2 q^5 t^9+2 t^7 q^6-4 t^8 q^5+t^9 q^4-t^7 q^5-t^8 q^4+3 t^5 q^6+2 t^6 q^5-2 t^7 q^4+t^8 q^3+t^4 q^6+2 t^5 q^5-4 t^6 q^4+2 t^7 q^3+3 t^4 q^5-3 t^5 q^4+t^7 q^2+3 t^3 q^5-t^4 q^4+2 t^6 q^2-5 t^4 q^3+2 t^5 q^2-4 t^3 q^3+2 q^2 t^4+2 t^5 q-t^2 q^3+t^3 q^2-q^3 t-3 t^2 q^2+2 t^3 q-q^2 t+2 t^2 q-t^2-q+t) + \ldots
$
\end{center}

\bibliography{dell-references}

\end{document}